\documentclass[11pt]{article}
\pdfoutput=1
\usepackage{jcapmod}
\usepackage{booktabs}
\usepackage[english]{babel}
\usepackage{amsmath,amssymb,amsbsy,amstext, amsthm, simplewick}
\usepackage{graphicx}
\usepackage{amsfonts}
\usepackage{amssymb}
\usepackage{upgreek}
 \usepackage{exscale,relsize}
 \usepackage[makeroom]{cancel}
\usepackage{soul}

\RequirePackage{color}

\newcommand{\rp}[1]{\textcolor{blue}{#1}}

\usepackage{colortbl}
\definecolor{green2}{cmyk}{0, 1, 0.5, 0}
\definecolor{lightgreen}{cmyk}{0.2, 0, 0.2, 0.2}
\definecolor{lightgray}{cmyk}{0.1,0.2,0,0.1}
\definecolor{lightgray2}{cmyk}{0.4,0.4,0,0.8}
\definecolor{black}{cmyk}{1.0,1.0,1.0,1.0}

\allowdisplaybreaks[1]

\usepackage{colortbl}
\definecolor{lightgreen}{cmyk}{0.2, 0, 0.2, 0.2}
\definecolor{lightgray}{cmyk}{0.1,0.2,0,0.1}
\definecolor{lightgray2}{cmyk}{0.1,0.1,0,0.1}

\makeatletter
\newlength{\apb@width}
\newcommand{\autoparbox}[2][c]{\settowidth{\apb@width}{#2}\parbox[#1]{\apb@width}{#2}}
\newcommand{\includegraphicsbox}[2][]{\autoparbox{\includegraphics[#1]{#2}}}
\makeatother

\numberwithin{equation}{section}

\def\beq{\begin{equation}}
\def\eeq{\end{equation}}

\def\bea{\begin{eqnarray}}
\def\eea{\end{eqnarray}}

\def\d{{\rm d}}

\def\beq{\begin{equation}}
\def\eeq{\end{equation}}
\def\bea{\begin{eqnarray}}
\def\eea{\end{eqnarray}}

\def\d{{\rm d}}

\def\Mp{M_{\rm pl}}
\def\d{{\rm d}}

\def\0{{\boldsymbol 0}}

\def\x{{\boldsymbol{x}}}

\def\fnl{f_{\mathsmaller{\rm NL}}}

\DeclareRobustCommand{\SkipTocEntry}[4]{}

\begin{document}

\begin{titlepage}

\setcounter{page}{1} \baselineskip=15.5pt \thispagestyle{empty}

\bigskip\

\vspace{1cm}
\begin{center}

{\fontsize{20}{28}\selectfont  \sffamily \bfseries B-modes and the Nature of Inflation}

\end{center}

\vspace{0.2cm}

\begin{center}
{\fontsize{13}{30}\selectfont  Daniel Baumann,$^{\bigstar}$ Daniel Green,$^{ \diamondsuit,\blacklozenge, \clubsuit}$ and Rafael A.~Porto$^{\spadesuit,\heartsuit }$}
\end{center}

\begin{center}

\vskip 8pt
\textsl{$^\bigstar$ D.A.M.T.P., Cambridge University, Cambridge, CB3 0WA, UK}
\vskip 7pt

\textsl{$^ \diamondsuit$ Canadian Institute for Theoretical Astrophysics, Toronto, ON M5S 3H8, Canada}
\vskip 7pt

\textsl{$^ \blacklozenge$ Stanford Institute for Theoretical Physics, Stanford University, Stanford, CA 94305, USA}
\vskip 7pt
\textsl{$^\clubsuit$ Kavli Institute for Particle Astrophysics and Cosmology, Stanford, CA 94305, USA}
\vskip 7pt
\textsl{$^\spadesuit$ Deutsches Elektronen-Synchrotron DESY, Theory Group, D-22603 Hamburg, Germany}
\vskip7pt
\textsl{$^\heartsuit$ Institute for Advanced Study, Princeton, NJ 08540, USA}\\
\end{center}

\vspace{1.2cm}
\hrule \vspace{0.3cm}
\noindent {\sffamily \bfseries Abstract} \\[0.1cm]
Observations of the cosmic microwave background do not yet determine whether inflation was driven by a slowly-rolling scalar field or involved another physical mechanism. In this paper we discuss the prospects of using the power spectra of scalar and tensor modes to probe the nature of inflation. We focus on the leading modification to the slow-roll dynamics, which entails a 
sound speed $c_s$ for the scalar fluctuations. We derive analytically a lower bound on $c_s$ in terms of a given tensor-to-scalar ratio $r$, taking into account the difference in the freeze-out times between the scalar and tensor modes. We find that any detection of primordial B-modes with $r > 0.01$ implies a lower bound on $c_s$ that is stronger than the bound derived from the absence of non-Gaussianity in the Planck data. 
For $r \gtrsim 0.1$, the bound would be tantalizingly close to a critical value for the sound speed, $(c_s)_\star = 0.47$ (corresponding to  $(\fnl^{\rm equil})_\star = -0.93$), which we show serves as a threshold for non-trivial dynamics beyond slow-roll. We also discuss how an order-one level of equilateral non-Gaussianity is a natural observational target for other extensions of the canonical paradigm.
\vskip 10pt
\hrule

\vspace{0.6cm}
 \end{titlepage}

 \tableofcontents

\newpage

\section{Introduction}

One of the central goals of modern cosmology is to determine the nature of inflation. While measurements of the cosmic microwave background (CMB) and the large-scale structure (LSS) are consistent with the predictions of single-field slow-roll models~\cite{PlanckInflation}, we should still ask to what degree observations 
 require that inflation occurred in this way. 
 
 \vskip 4pt
 A systematic way to describe deformations of the canonical framework is the effective field theory (EFT) of inflation~\cite{Cheung:2007st}.
 This is a theory of the two massless fields that are guaranteed to be present in any model of inflation: the Goldstone boson of spontaneously broken time translations, $\pi$, and the graviton, $h_{ij}$.  In single-field slow-roll inflation the role of the Goldstone boson is played by fluctuations in the inflaton field, which satisfy a relativistic dispersion relation, $\omega = k$, and are only very weakly interacting. Deviations from slow-roll inflation are parameterized by a non-trivial dispersion relation $\omega(k)$,
 higher-order self-interactions of $\pi$ or couplings to other fields. A significant advantage of the EFT framework is that it allows us to study scenarios where the accelerated expansion is not necessarily driven by a weakly coupled fundamental scalar~field.

A well-motivated possibility is to modify the Goldstone dispersion relation by adding a sound speed, $\omega = c_s k$ \cite{ArmendarizPicon:1999rj}.   While perturbative higher-derivative corrections to the slow-roll dynamics can only induce small deviations from $c_s =1$ \cite{Creminelli:2003iq},
models with $c_s \ll 1$ are characteristic of non-perturbative physics~\cite{Silverstein:2003hf} or non-trivial dynamics~\cite{Baumann:2011su}. 
The difference between weakly and strongly coupled inflationary backgrounds is an important qualitative distinction.
As we will show, for the theory to be weakly coupled 
at all relevant energies, the sound speed has to be above the critical value 
 \beq 
\label{eq:th} (c_s)_\star = 0.47\ .
\eeq 
Finding that $c_s  > (c_s)_\star$, would therefore be a strong indication in favor of the standard scenario, while $c_s  < (c_s)_\star$ requires physics beyond the slow-roll paradigm. 

\begin{figure}[htbp]
\begin{center}
\includegraphics[width=0.85\textwidth]{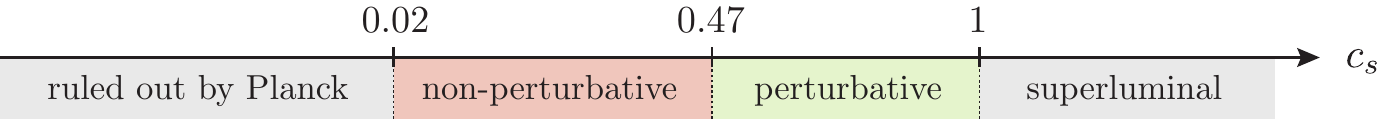}
\caption{Graphical illustration of the threshold value $(c_s)_\star =0.47$.}
\label{SoundAxis}
\end{center}
\end{figure}

There are several avenues for probing a sound speed of the inflationary perturbations. Famously, small $c_s$ enhances scalar self-interactions and hence leads to non-Gaussian correlations. The amplitude of the induced equilateral bispectrum is $\fnl^{\rm equil} \propto c_s^{-2}$, and the absence of significant non-Gaussianity in the Planck data~\cite{PlanckNG} puts a lower bound on the allowed value of the sound speed: $c_s > 0.02~({\rm 95\%CL}).$ As we shall see, the critical sound speed~(\ref{eq:th}) corresponds to equilateral non-Gaussianity with $(f_{\mathsmaller{\rm NL}}^{\rm equil})_\star  = -0.93$, which is two orders of magnitude below the sensitivity of Planck~\cite{PlanckNG} but two orders of magnitude above the slow-roll expectation. 
This result is surprisingly robust, as similar thresholds, with $|f_{\mathsmaller{\rm NL}}^{\rm equil}| \simeq {\cal O}(1)$, exist for other cubic Goldstone interactions even when $c_s =1$.  It is a universal feature that order-one equilateral non-Gaussianity separates qualitatively distinct regions in the parameter space of the EFT~of~single-field inflation~\cite{Cheung:2007st} and extensions thereof~\cite{Senatore:2010wk, LopezNacir:2011kk, Baumann:2011nk}.
 Probing equilateral non-Gaussianity down to the order-one level is therefore an important and well-motivated observational target~\cite{Abazajian:2013vfg, LeoTalk, DanTalk, RafaelTalk,Baumann:2014nda}.
\vskip 4pt
The BICEP2 collaboration~\cite{Ade:2014xna} has recently set a new standard for measurements of B-mode polarization in the CMB by reporting a more than $5\sigma$ detection at degree angular scales. 
While we await confirmation of the primordial origin of the BICEP2 signal,
 it is a timely issue to address the implications of a detectable tensor-to-scalar ratio for the physics of inflation. In this paper we will show that any detection of primordial tensor modes with $r>0.01$ puts a lower bound on~$c_s$ that is stronger than the Planck-only constraint.

Our bound originates from the fact that a small value of $c_s$ boosts the amplitude of scalar fluctuations and hence suppresses the tensor-to-scalar ratio, which at leading order is given by $r =  16 \varepsilon_1 c_s$,
 where $\varepsilon_1 \equiv - \dot H/H^2$ quantifies the deviation of the background from perfect de Sitter space. 
 A large value of $r$ is only compatible with small $c_s$ if the value of $\varepsilon_1$ is much larger than in the canonical expectation in slow-roll models~\cite{Creminelli:2014oaa}: e.g.~$\varepsilon_1 \simeq 0.01$ in $m^2 \phi^2$ chaotic inflation.  However, for $\varepsilon_1 \gtrsim 0.1$, the expression for $r$ receives an important correction due to the fact that scalar and tensor fluctuations freeze out at different times, 
\beq
  \label{eq:TS2}
r = 16 \varepsilon_1 c_s \left(\frac{H_t}{H_s}\right)^2\ , 
\eeq
where $H_t$ and $H_s$ denote the Hubble scales at the two freeze-out times. 
 Since $\dot H < 0$, when the null energy condition is satisfied, we have $H_t < H_s$ and the tensor-to-scalar ratio receives an additional suppression. This suppression becomes more significant for larger values of $\varepsilon_1$ (faster evolution of $H$) and smaller values of $c_s$ (larger separation of the freeze-out times). This feature, together with constraints from the shape of the scalar power spectrum, leads to an analytic lower bound on $c_s$ for a given tensor-to-scalar ratio. The ultimate expression for the bound depends on various factors which we will explain in detail throughout the paper. The upshot, however, is clear:  A large B-mode signal constrains the sound speed much better than measurements of CMB temperature fluctuations alone. Specifically, for $r > 0.13$ --- corresponding to the level reported by BICEP2 and also the benchmark of $m^2\phi^2$ inflation --- we find\footnote{In order to arrive at \eqref{eq:cs1} we have assumed a constant sound speed. This is the case that is most relevant for a direct comparison with the Planck limit $c_s > 0.02~(95\%{\rm CL})$. We will also present bounds on $c_s$ that do not make this assumption yet lead to qualitatively similar results.}
\beq
 c_s \,>\, 0.14 \ ,  \label{eq:cs1} 
\eeq
which would correspond to a bound on the $c_s$-induced non-Gaussianity of order $|\fnl^{\rm equil}| \lesssim 10$. Notice that this would be an order of magnitude stronger than the previous Planck-only bound: $\fnl^{\rm equil} = -42 \pm 75$ \cite{PlanckNG}.

To test our analytical approach we perform a joint likelihood analysis of the CMB data from WMAP, Planck and BICEP2 {\it without} foreground subtraction.\footnote{We warn the reader this analysis is performed as a case of study only, and future maps will be required to properly assess the level of foregrounds in the BICEP2 region, e.g.~\cite{Adam:2014bub,Flauger:2014qra, Mortonson:2014bja}. Fortunately, the debate will soon be settled by future CMB polarization measurements.} We find, for constant sound speed, $c_s \,>\, 0.25\ \ (95\%{\rm CL})$, which is consistent with \eqref{eq:cs1} and supports our analytic arguments. (The small discrepancy is almost entirely due to the preference for a negative running of the spectral index in the CMB data, which favors a larger suppression factor in \eqref{eq:TS2}. The tendency for negative running disappears if the low-$\ell$ data is removed, and the results from the numerical and analytic studies approach each other.)

Although a detection of primordial B-modes with $r \gtrsim 0.1$ would lead to a lower bound on $c_s$ remarkably close to the threshold value in \eqref{eq:th}, this does not mean non-Gaussianity may be absent in future observations~\cite{Creminelli:2014oaa,  MatiasTalk, DAmico:2014cya, LeoTalk}. In fact, the bispectrum remains an important observable, and there exist at least three distinct possibilities for a detectable signal:  First of all, values of $c_s$ and $\fnl^{\rm equil}$ near the threshold (i.e.~$c_s \simeq 0.5$ and $|f_{\mathsmaller{\rm NL}}^{\rm equil}| \simeq {\cal O}(1)$) arise naturally from strongly coupled backgrounds with a single scale. Second, there are other (technically natural) deformations of slow-roll inflation that are unrelated to the sound speed~\cite{Cheung:2007st,Senatore:2009gt} and therefore allow for potentially observable equilateral non-Gaussianity irrespective of a bound.  Finally, the presence of additional light degrees of freedom during inflation may also induce a detectable bispectrum. Some of these scenarios, such as quasi-single-field inflation~\cite{Chen:2009zp} or non-adiabatic dissipative effects~\cite{Berera:1995ie, LopezNacir:2011kk}, can lead to sizable equilateral non-Gaussianity. These cases may be distinguished from the first two options through correlations between the equilateral shape and certain squeezed limits.

\vskip 10pt
The outline of the paper is as follows: In Section~\ref{sec:bound}, we derive an analytic bound on the sound speed for a given value of the tensor-to-scalar ratio. This result follows from the expression in \eqref{eq:TS2} and relies on minimal input from the data. 
In  Section~\ref{sec:data}, we present a joint analysis of data from WMAP, Planck and BICEP2 (without foreground subtraction) which confirms our analytic expectation. 
We discuss the robustness of our results to variations in the inflationary parameters and changes in the cosmological data sets. In Section~\ref{sec:threshold}, we obtain a critical value for $c_s$, and a related threshold for $\fnl^{\rm equil}$, from a unitarity bound in the EFT of inflation. The threshold divides perturbative slow-roll inflation from strongly coupled backgrounds for sound speed models. In Section~\ref{sec:NG}, we survey a range of well-motivated scenarios, 
consistent with current data, which may produce a detectable level of equilateral non-Gaussianity even for $c_s$ close to 1.
 We conclude with a summary and an outlook on future prospects in Section~\ref{sec:conclusions}.

\section{Implications of a B-mode Detection}
\label{sec:bound}

We begin with an analytic discussion of the consequences of a B-mode detection for inflationary models with a sound speed (see~also~\cite{Creminelli:2014oaa, MatiasTalk, LeoTalk, DAmico:2014cya}).  The main result of this section will be a 
lower bound on the sound speed as a function of the tensor-to-scalar ratio. 

\subsection{Spectra of Primordial Perturbations}

Quantum fluctuations of any massless fields get amplified during inflation.
Two massless fields that are guaranteed to exist in any model of inflation are the curvature perturbation $\zeta$ and the graviton $h_{ij}$. In comoving gauge, these fields are isotropic and anisotropic perturbations to the spatial metric, respectively,
\beq
g_{ij}(t,\x) = a^2(t) \left[e^{2\zeta(t,\x)}\delta_{ij} + 2h_{ij}(t,\x) \right]\ ,
\eeq
where $h_{ij}$ is transverse and traceless. The scale factor $a(t)$ is that of an arbitrary quasi-de Sitter background with Hubble expansion rate $H(t) \equiv \partial_t \ln a$.  The dynamics of $\zeta$ may contain a time-dependent sound speed $c_s(t)$.
To describe the evolution of the Hubble parameter and the sound speed, it is convenient to introduce the following {\it flow parameters}\hskip 2pt\footnote{In an abuse of terminology, we will sometimes refer the Hubble flow parameters $\varepsilon_n$ and the sound flow parameters~$\delta_n$ as the `slow-roll parameters'.}
 \begin{align}
\varepsilon_{n+1} &\equiv \frac{d \ln \varepsilon_n}{dN}\ , \qquad \varepsilon_0 \equiv \frac{H(N_i)}{H(N)} \label{equ:epsn} \ , \\
\delta_{n+1} &\equiv \frac{d \ln \delta_n}{dN} \ , \qquad \delta_0 \equiv \frac{c_s(N)}{c_s(N_i)}\ , \label{equ:deltan}
\end{align}
where $N \equiv \ln a$ and $N_i$ denotes an arbitrary reference time. Provided fluctuations are produced in the vacuum,\footnote{Scalar and tensor perturbations may also be produced from non-vacuum states. This possibility was studied in detail in \cite{LopezNacir:2011kk, Senatore:2011sp, Cook:2011hg}, where it was shown that non-Gaussianity may be able to discern these types of models. } the power spectra of $\zeta$ and $h_{ij}$ in models with a sound speed are
\begin{align}
\Delta_\zeta^2(k) &\,=\, \frac{1}{8\pi^2} \frac{H^2}{\Mp^2} \frac{1}{\varepsilon_1 c_s}  \Bigg|_{c_sk=aH} \ , \label{equ:Dz} \\[4pt]
\Delta_h^2(k) &\,=\, \frac{2}{\pi^2} \frac{H^2}{\Mp^2} \Bigg|_{k=aH} \label{equ:Dh} \ ,
\end{align}
to leading order in the flow parameters. Notice that while the scalar fluctuations are evaluated when modes cross the sound horizon, $c_sk=aH$, the tensor fluctuations are evaluated at $k=aH$.  This fact will play a key role in what follows. As long as the slow-roll conditions are satisfied, $|\varepsilon_n| \ll 1$ and  $|\delta_n| \ll 1$, the induced scalar spectrum has an approximate power-law form 
 \beq
 \Delta_\zeta^2 = A_s\left(\frac{k}{k_0}\right)^{n_s - 1 + \frac{1}{2} \alpha_s \ln(k/k_0)}\ ,
 \eeq 
 with $A_s \simeq 2.2 \times 10^{-9}$ and, to leading order, 
 \begin{align}
n_s - 1 &= -2 \varepsilon_1 - \varepsilon_2 - \delta_1 \ , \label{eq:ns} \\
\alpha_s &= - 2\varepsilon_1  \varepsilon_2  -  \varepsilon_2 \varepsilon_3 - \delta_1 \delta_2\ .\label{eq:alphas}
\end{align}
The right-hand sides in (\ref{eq:ns}) and (\ref{eq:alphas}) are to be evaluated when the pivot scale $k_0$ crosses the sound horizon. 
Notice that the scalar running $\alpha_s$ starts at second order in the flow parameters. This is a key feature of inflationary models, which will be important in our analysis below.
Throughout this section, we will assume $\varepsilon_3 \ll \varepsilon_{1,2}$ and $\delta_2 \ll \delta_1$. 
This approximation will be relaxed in Section~\ref{sec:data}, where we present a numerical analysis at higher order in the slow-roll expansion.

\subsection{Origin of a Bound on the Sound Speed}

\begin{figure}[h!]
\begin{center}
\includegraphics[width=0.65\textwidth]{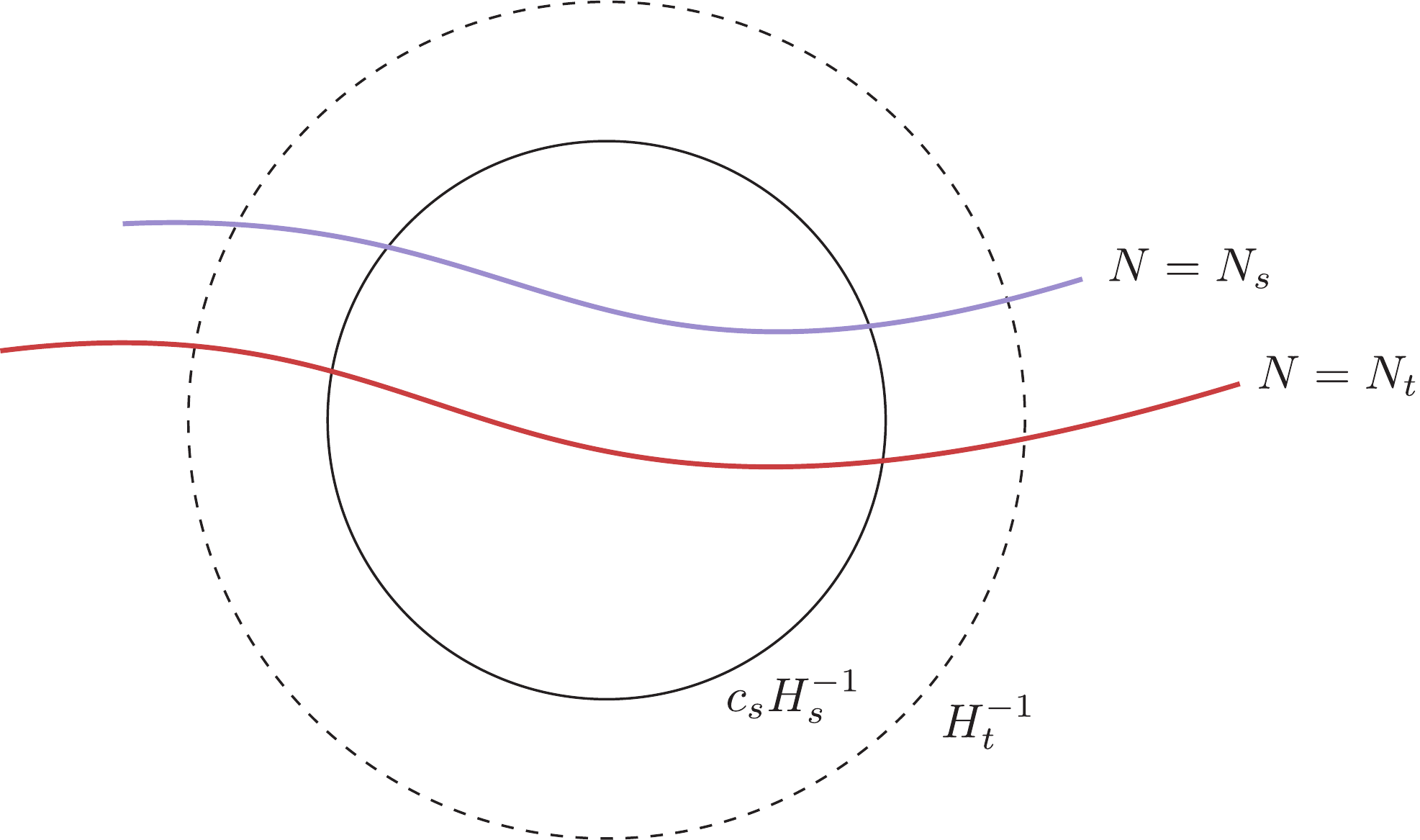} 
\caption{For $c_s < 1$, scalars and tensors freeze out at different times. }
\label{fig:freeze}
\end{center}
\end{figure}

The fact that tensors and scalars freeze out at different times is important, as it implies that the tensor-to-scalar ratio will depend on the ratio of the Hubble scales at the two freeze-out times:
\beq
r \equiv \frac{\Delta_h^2}{\Delta_\zeta^2} \,=\, 16 \varepsilon_1 c_s \left(\frac{H_t}{H_s}\right)^2\ , \label{equ:r}
\eeq
where $H_t \equiv H(N_t)$ and $H_s \equiv H(N_s)$ denote the Hubble parameters at $k=aH$ and $c_s k= aH$, respectively. 
For $c_s < 1$, the sound horizon is smaller than the Hubble radius and the scalars freeze out before the tensors do  (see fig.~\ref{fig:freeze}). 
Since $\dot H < 0$ by virtue of the null energy condition, we have $H_t / H_s < 1$ and the tensor-to-scalar ratio in (\ref{equ:r}) receives an additional suppression.  
In order to quantify this effect, we evolve $H(t)$ in the slow-roll approximation.
At next-to-leading order, we have
\beq
H_t = H_s(1 - \varepsilon_1 (N_t-N_s)+ \cdots)  \simeq H_s (1+\varepsilon_1 \ln c_s + \cdots) \ ,
\eeq
where $\varepsilon_1 \equiv \varepsilon_1(N_s)$ and the ellipses denote terms at higher order in $\varepsilon_1$ and $\varepsilon_{n \ge 2}$.
Hence, we get
\beq
r  \, \simeq\, 16 \varepsilon_1 c_s \Big[1+2\varepsilon_1 \ln c_s + \cdots \Big]\ . \label{equ:r2}
\eeq
The leading order expression, $r=16 \varepsilon_1 c_s$, implies that $r\simeq 0.1$ with $c_s \simeq 0.02$ requires $\varepsilon_1 \simeq 0.3$. However, for such a large value of $\varepsilon_1$ and such a small value of $c_s$, the correction in (\ref{equ:r2}) is not small, i.e. $2\varepsilon_1 \ln c_s \simeq -2.4$.
To treat this regime of the parameter space, we need to understand the $\ln c_s$-enhanced contributions in the slow-roll expansion.

By definition, the ratio $H_t/H_s$ is determined by an integral over the time-dependent Hubble flow parameter~$\varepsilon_1(N)$:
\beq
\ln (H_t/H_s) = - \int_{N_s}^{N_t} \varepsilon_1(N) \, \d N \,\equiv\, - \bar\varepsilon_1 \Delta N\ , \label{equ:H}
\eeq
where $\Delta N \equiv N_t - N_s$ and $\bar\varepsilon_1 \equiv \varepsilon_1(\bar N)$, for $\bar N \in [ N_s, N_t]$ as guaranteed by the mean value theorem. 
At the same time, the horizon crossing conditions $k=aH$ and $c_sk = a H$ imply
\beq
\frac{H_t}{H_s} = \frac{1}{c_s} \frac{a_s}{a_t}  = \frac{1}{c_s} e^{-\Delta N} \ . \label{equ:cross}
\eeq
Combining (\ref{equ:H}) and (\ref{equ:cross}), we find
\beq
\label{equ:DNapp}
\Delta N = \frac{-\ln c_s}{1-\bar \varepsilon_1}\ ,
\eeq
which, after substitution into (\ref{equ:H}) and \eqref{equ:r}, leads to
\beq
r = 16 \varepsilon_1 c_s^{\frac{1+\bar\varepsilon_1}{1-\bar\varepsilon_1}}\ . \label{equ:r3}
\eeq
 We see that the tensor-to-scalar ratio is suppressed for $\bar \varepsilon_1 \gtrsim \varepsilon_1$, while this effect can be avoided if the dynamics leads to $\bar \varepsilon_1 \ll \varepsilon_1$.  In order for (\ref{equ:r3}) to become a meaningful expression, we need to determine the auxiliary parameter $\bar\varepsilon_1$ in terms of the Hubble flow parameters $\varepsilon_n$ evaluated at~$N_s$.  

\begin{figure}[htbp]
\begin{center}
\includegraphics[scale=0.4]{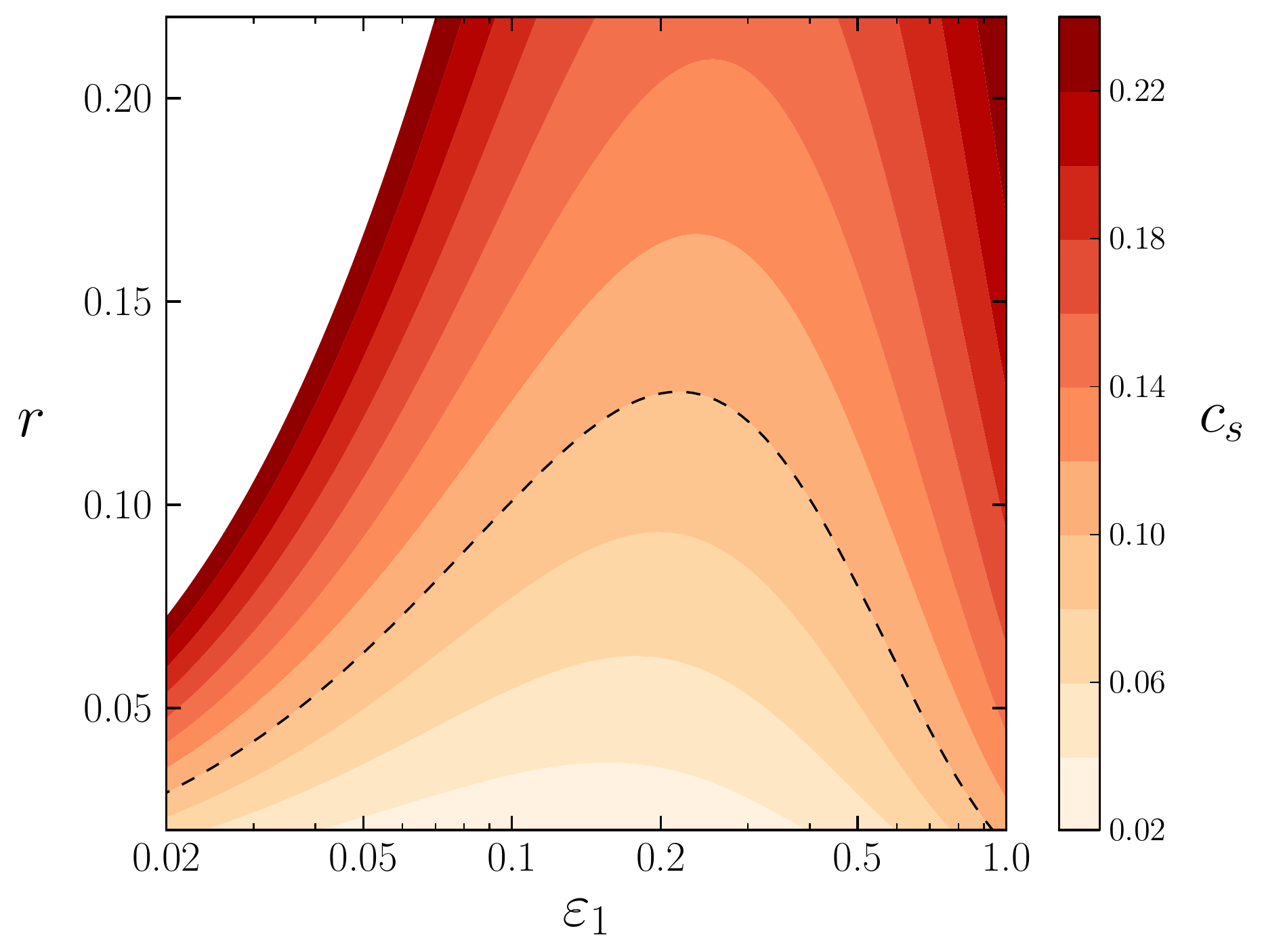} \hskip 20pt
\includegraphics[scale=0.4]{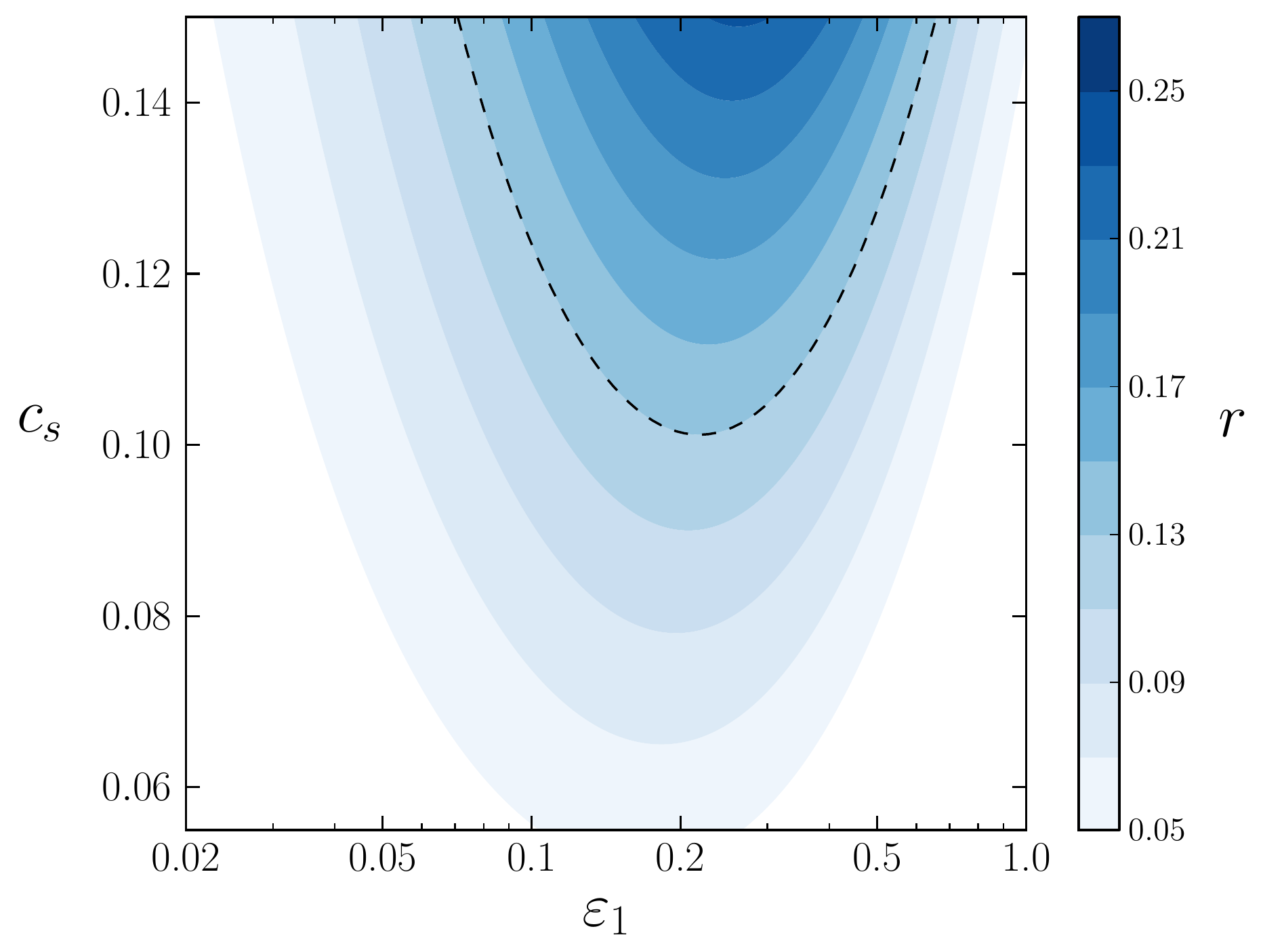}
\vskip -8pt
\caption{Plots of $c_s(r,\varepsilon_1)$ ({\it left}) and $r(c_s,\varepsilon_1)$ ({\it right}) as given by~(\ref{equ:r4}). The dashed contour in the left plot shows the value of $r$ for $c_s =0.1$. It has a maximum of $r=0.13$ for $\varepsilon_1 = 0.2$. The dashed contour in the right plot illustrates the lower bound on $c_s$ for $r=0.13$.}
\label{rcsep}
\end{center}
\end{figure}

\vskip 4pt
To leading order in the slow-roll expansion, we have $\varepsilon_1 \simeq const.$ and hence $\bar \varepsilon_1 \simeq \varepsilon_1$.
Since we are not keeping higher orders (unless they are $\ln c_s$-enhanced), we should expand the exponent in (\ref{equ:r3}):
\beq
r \simeq 16 \varepsilon_1 c_s^{1+2\varepsilon_1}\ . \label{equ:r4}
\eeq
This result corresponds to a resummation of the {\it leading logarithm} of (\ref{equ:r2}), i.e.~it is valid to all orders in $\varepsilon_1 \ln c_s$, but only holds to leading order in $\varepsilon_1$. As expected, $r$ in (\ref{equ:r4}) receives an additional suppression for large $\varepsilon_1$ and small $c_s$. In fact, at fixed $c_s$, the tensor-to-scalar ratio reaches a maximum value at $\varepsilon^{\rm max}_1 = - (2\ln c_s)^{-1}$, as can be seen in the left panel of fig.~\ref{rcsep}.
Turning this around, a detection of $r$ would imply a lower bound on the sound speed, as shown in the right panel of fig.~\ref{rcsep}.
We see that $r > 0.13$ requires
\beq
c_s > 0.1\ . \label{equ:bound}
\eeq
Notice that even values of $r$ an order of magnitude smaller would give a bound on $c_s$ that is as strong as the bound derived from the Planck measurement of the bispectrum.

\begin{figure}[htbp]
\begin{center}
\includegraphics[scale=0.4]{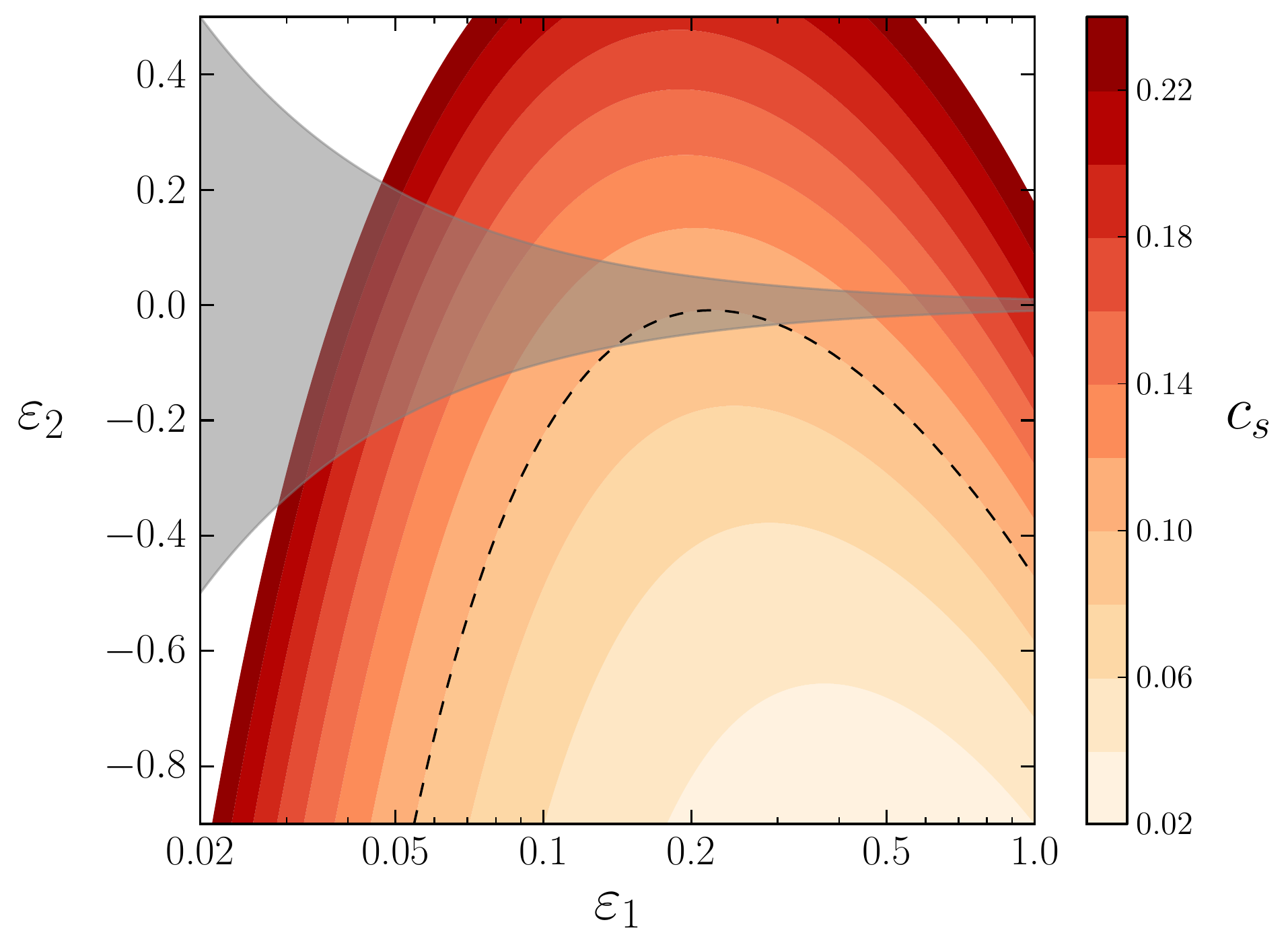} \hskip 20pt
\includegraphics[scale=0.4]{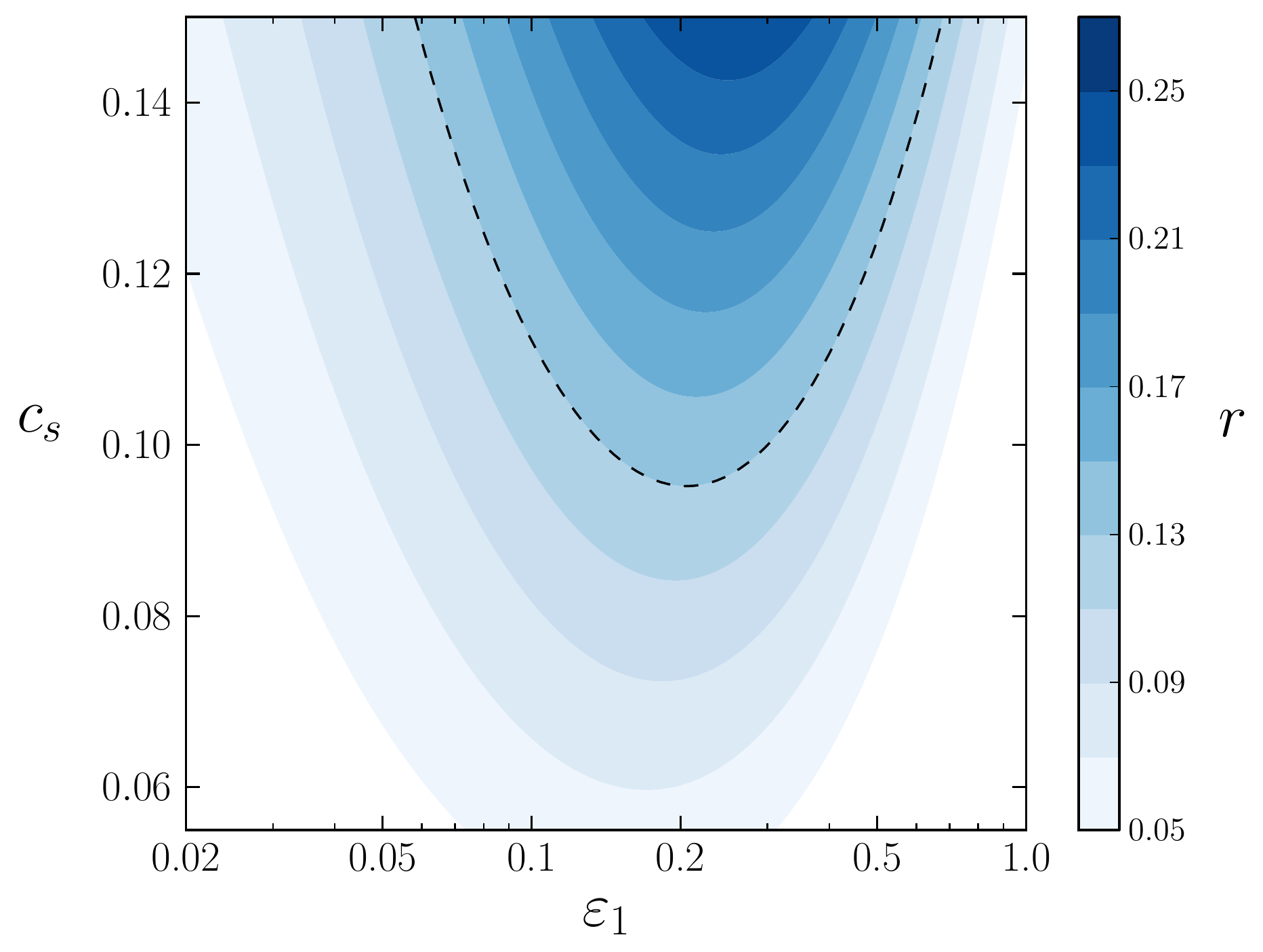}
\vskip -8pt
\caption{ {\it Left:} plot of $c_s(\varepsilon_1, \varepsilon_2)$, as given by~(\ref{equ:r5}), for $r = 0.13$. The dashed contour corresponds to the previous bound $c_s=0.1$. The grey shaded area is the region that is consistent with the constraint from the running: $| \varepsilon_1 \varepsilon_2| < 0.01$.  {\it Right:} plot of $c_s(r,\varepsilon_1,\varepsilon_2)$ as given by~(\ref{equ:r5}) with $\varepsilon_2 \to -0.01/\varepsilon_1$ (the most negative value consistent with the bound on $\alpha_s$).}
\label{fig:rcs2}
\end{center}
\end{figure}

We emphasize that the bound in (\ref{equ:bound}) is model-dependent and has assumed that $\varepsilon_1(N)$ does not vary significantly. 
As we already alluded to, the lower bound on $c_s$ may be relaxed if $\bar \varepsilon_1 \ll \varepsilon_1$ in (\ref{equ:r3}). This is achieved for large and negative values of~$\varepsilon_2$,\footnote{The case $\bar\varepsilon_1 > \varepsilon_1$, namely $\varepsilon_2 >0$, obviously introduces extra suppression in $r$, which strengthens the bound.}  so that $\varepsilon_1$ is large at sound horizon crossing, but decreases quickly thereafter. However, a non-negligible $\varepsilon_2 $ also produces large logarithms. 
In Appendix~\ref{sec:flow}, we show how to resum the $\varepsilon_2 \ln c_s$ terms. The result is $\bar\varepsilon_1 = \varepsilon_1 (1-c_s^{-\varepsilon_2})/(\varepsilon_2 \ln c_s)$, which implies
\beq
r = 16 \varepsilon_1 c_s^{1+2 \varepsilon_1\cdot (1-c_s^{-\varepsilon_2})/(\varepsilon_2 \ln c_s)} \ . \label{equ:r5}
\eeq
 In the left panel of fig.~\ref{fig:rcs2}, we see that $r \simeq 0.13$ is now indeed consistent with small $c_s$ if both $\varepsilon_1$ and $|\varepsilon_2|$ are large. However, having both of these parameters take on large values induces a significant running of the spectral index $\alpha_s$, c.f.~\eqref{eq:alphas}. Currents constraints on the running thus restrict the size of $\varepsilon_1 \varepsilon_2$. Indeed, imposing $|\alpha_s|  \lesssim 2 \times10^{-2}$ \cite{PlanckInflation},\footnote{This value follows from the 95\% confidence interval, while taking $\alpha_s=0$ as the central value. We will take into account the full data set in Section \ref{sec:data}.} we recover a bound on $c_s$ that is only marginally weaker than the previous bound; see the right panel of fig.~\ref{fig:rcs2}.

\subsection{Degeneracies and Second-Order Corrections}
\label{sec:degeneracies}

Up until now we have only used minimal data input to derive an analytic constraint on $c_s$: a~lower bound on $r$ and an upper bound on $|\alpha_s|$. When we perform a detailed CMB analysis, in the next section, we will get additional constraints from the precise shape of the scalar spectrum. In particular, we have yet to take into account its near scale-invariance, i.e.~$n_s \simeq 0.96$ \cite{PlanckInflation}. Moreover, the numerical value of the bound on $c_s$ will be affected by second-order corrections to the primordial spectra (\ref{equ:Dz}) and (\ref{equ:Dh}), which so far we have not incorporated. 
We conclude this section with a brief discussion on how the main feature of our complete analysis --- a stronger  bound on $c_s$ --- can also be understood analytically.

\vskip 4pt
The most important second-order effect is a correction to the expression \eqref{equ:r5} for the tensor-to-scalar ratio
\beq
\label{bracket}
r = 16 \varepsilon_1 c_s^{1+2 \varepsilon_1\cdot (1-c_s^{-\varepsilon_2})/(\varepsilon_2 \ln c_s)}  \Big[1-{\cal C} \varepsilon_2 +(2-{\cal C})\delta_1\Big]\ ,
\eeq
where ${\cal C} \simeq 0.73$. The importance of the extra terms in the square brackets depends on the sizes of the flow parameters $\varepsilon_2$ and $\delta_1$.
Since both parameters appear at leading order in the expression for $n_s$, c.f.~(\ref{eq:ns}), they are constrained by the near scale-invariance of the spectrum.
Moreover, we have seen that the combination of large $r$ and small $c_s$ typically requires relatively large values of~$\varepsilon_{1}$. Consistency with the measured spectral index then means that either $\delta_1$ or $\varepsilon_2$ (or a combination thereof) has to be chosen to cancel part of the contribution from $\varepsilon_1$.
  These two possibilities correspond to two perfect degeneracies that keep $n_s$ within experimental bounds: 
  \begin{itemize}

\item ${\boldsymbol{ \delta_1 \simeq -2 \varepsilon_1}}$\\
Allowing for a time dependence in the sound speed gives enough freedom to satisfy the constraints from $n_s$ and $\alpha_s$ simultaneously. In particular, the time variation of~$c_s(t)$ can be chosen to nearly cancel the effect of the rather rapid evolution of the Hubble parameter~$H(t)$. Setting $\delta_1 \simeq - 2 \varepsilon_1$ makes any value of $\varepsilon_1$ consistent with the measured value of~$n_s$. Moreover, if $|\varepsilon_2| \lesssim 10^{-2}$ then $\varepsilon_1$ is not constrained by the bound on the running~$\alpha_s$. The analysis is then similar to what led us to fig.~\ref{rcsep}. However, this time we include the second-order correction in $r$, which along the degeneracy $\delta_1 \simeq - 2 \varepsilon_1$ becomes 
\beq
r \simeq 16 \varepsilon_1 c_s^{1+2 \varepsilon_1}  \Big[1-2.54\hskip 1pt \varepsilon_1\Big]\ .
\eeq
Since $\varepsilon_1 > 0$, the correction in the square brackets suppresses $r$ and we expect a stronger bound on $c_s$ compared to what we found in \eqref{equ:bound}.
Indeed, $r > 0.13$ now requires  
\beq
c_s > 0.15 \quad\quad {\rm for}\ \ \delta_1 \simeq - 2\varepsilon_1 \ . \label{eq:cs1n}
\eeq
\item ${\boldsymbol{ \varepsilon_2 \simeq -2 \varepsilon_1}}$\\ 
Allowing for a varying $\varepsilon_1$, a potentially large contribution to $n_s$ from $\varepsilon_1$ may be cancelled by choosing $\varepsilon_2 \simeq - 2 \varepsilon_1$ while keeping $c_s$ constant (i.e.~$\delta_1=0$). Along this degeneracy curve any value of $\varepsilon_1$ is consistent with the observed spectral index. 
However, the running becomes $\alpha_s = - 2\varepsilon_1 \varepsilon_2 \to 4 \varepsilon_1^2$, so that the constraint $|\alpha_s| < 2\times 10^{-2}$ implies 
 \beq
 \varepsilon_1 < 0.07\ . \label{equ:as}
 \eeq
Moreover, the tensor-to-scalar ratio along the degeneracy is given by
\beq
r \,\simeq\, 16 \varepsilon_1 c_s^{1 - (1-c_s^{2\varepsilon_1})/ \ln c_s}  \Big[1+1.46\hskip 1pt \varepsilon_1\Big]\ . \label{equ:r1}
\eeq
Combining \eqref{equ:r1} and \eqref{equ:as}, and imposing $r > 0.13$, then leads to 
\beq
c_s > 0.14 \quad\quad {\rm for}\ \ \varepsilon_2 \simeq -2\varepsilon_1 \ . \label{eq:cs0n}
\eeq
Notice that \eqref{eq:cs0n} can be 
read off from fig.~\ref{fig:rcs2} as the overlap between $\varepsilon_1 \simeq 0.07$ and the grey shaded region. 

\end{itemize}
In the next section, we will present a complete CMB analysis. Remarkably, we will find only small departures from the bound we arrived at analytically.

\section{CMB Analysis} 
\label{sec:data}

A complete likelihood analysis can differ from the analytic bounds of the previous section because of additional degeneracies between parameters or stronger constraints arising from the precise form of the scalar power spectrum.  
We would like to understand to what degree our analytic estimates have accounted for these effects.  For purposes of illustration, we will perform a joint likelihood analysis of data from WMAP, Planck, and BICEP2. 
We appreciate the significant level of uncertainty in the modelling of dust foregrounds in the BICEP2 region of the sky~\rp{\cite{Adam:2014bub, Flauger:2014qra, Mortonson:2014bja}}.  Our analysis will use the BICEP2 likelihood without foreground subtraction, but we caution the reader that quantitative details of our results are subject to change in the event that the BICEP2 likelihood is significantly revised after a better understanding of the foregrounds.
Our goal in this section therefore isn't to derive a quantitative bound on the sound speed, but to understand if our analytic approach could have missed an unforeseen degeneracy that would allow the bound on $c_s$ to be evaded completely. We will find that this is not the case and our analytic result is therefore a good reflection of what could be expected of future data analyses.
A similar analysis for the case of slow-roll inflation has recently appeared in~\cite{Martin:2014lra}.

\subsection{Inflationary Spectra to Second Order}
\label{sec:2nd}

For consistency, we will use results for the scalar and tensor power spectra at second order in the Hubble flow parameters (\ref{equ:epsn}) and the sound flow parameters (\ref{equ:deltan}); see~\cite{Chen:2006nt, Lorenz:2008et, Agarwal:2008ah, Powell:2008bi, Martin:2013uma}.
We write the power spectra in the standard power-law form 
\begin{align}
\Delta_\zeta^2(k) &=  A_s \left(\frac{k}{k_0}\right)^{n_s - 1 + \frac{1}{2} \alpha_s \ln(k/k_0)}\ ,
 \label{equ:scalar} \\[2pt]
\Delta_h^2(k) &= r A_s \left(\frac{k}{k_0}\right)^{n_t + \frac{1}{2} \alpha_t \ln(k/k_0)}\ , \label{equ:tensor}
\end{align}
where $k_0 =0.05\, {\rm Mpc}^{-1}$ is our chosen pivot scale.
At second order, we then have
\begin{align}
n_s - 1 &\,=\, - 2 \varepsilon_1 - \varepsilon_2 - \delta_1  \nonumber\\
&\hspace{0.7cm} -2\varepsilon_1^2 - (3-2{\cal C}) \varepsilon_1 \varepsilon_2 + {\cal C}\hskip 1pt \varepsilon_2 \varepsilon_3 - 3 \varepsilon_1\delta_1 -\varepsilon_2\delta_1 -(2-{\cal C}) \delta_1\delta_2 - \delta_1^2 \ , \label{equ:ns} \\[6pt]
\alpha_s &\,=\,  - 2\varepsilon_1 \varepsilon_2 - \varepsilon_2 \varepsilon_3 - \delta_1 \delta_2 \label{equ:alphas}\ ,\\[6pt]
n_t  &\,=\, - 2 \varepsilon_1 - 2 \varepsilon_1^2  - 2(1-{\cal C}-\ln c_s) \varepsilon_1 \varepsilon_2  \ , \\[6pt]
\alpha_t &\,=\,  -2\varepsilon_1 \varepsilon_2\ , \\[6pt]
r &\,=\, 16 \varepsilon_1 c_s  \Big[1+2 \varepsilon_1 \ln c_s-{\cal C} \varepsilon_2 +(2-{\cal C})\delta_1\Big]\ ,
\end{align}
where ${\cal C} \equiv 2 - \ln 2 - \gamma$ (with $\gamma = 0.5772$ the Euler-Mascheroni constant). Even the indices $n_t$ and $\alpha_t$ are now expressed in terms of slow-roll parameters evaluated when the pivot scale crosses the sound horizon, $c_s k_0 = aH$.  This is responsible for the $\ln c_s$-terms in the above formulas.  In Appendix~\ref{sec:flow}, we show that the resummation of these logarithms gives
\beq
r = 16 \varepsilon_1 c_s^{1+2 \varepsilon_1\cdot (1-c_s^{-\varepsilon_2})/(\varepsilon_2 \ln c_s)}  \Big[1-{\cal C} \varepsilon_2 +(2-{\cal C})\delta_1\Big]\ .  \label{equ:RR}
\eeq
We will use this expression for the tensor-to-scalar ratio in our analysis.
This allows us to explore smaller values of $c_s$ while maintaining perturbative control. 

\subsection{Joint Analysis of Planck and BICEP2}
\label{sec:joint}

This section describes the methodology and the results of a dedicated likelihood analysis using data from WMAP, Planck and BICEP2. 

\vskip 4pt
The Planck likelihoods were described in detail in \cite{Ade:2013kta}.
We use the {\tt CamSpec} likelihood for the Planck temperature power spectrum in the multiplole range $50 \le \ell \le 2500$ and the {\tt Commander} likelihood for $2 \le \ell \le 49$.  As in the Planck analysis~\cite{Ade:2013zuv}, we use the WMAP polarization data for $\ell \le 32$.  We will also show results with the low-$\ell$ likelihoods excluded. This helps to identify effects that are driven by the low-$\ell$ anomalies.  We use all nine bandpowers of the BICEP2 observations~\cite{Ade:2014xna,Ade:2014gua}. Pending a resolution in the debate about the importance of dust foregrounds in the BICEP2 region, we use the BICEP2 likelihood without foreground subtraction.

 \begin{table}[h!]

	\heavyrulewidth=.08em
	\lightrulewidth=.05em
	\cmidrulewidth=.03em
	\belowrulesep=.65ex
	\belowbottomsep=0pt
	\aboverulesep=.4ex
	\abovetopsep=0pt
	\cmidrulesep=\doublerulesep
	\cmidrulekern=.5em
	\defaultaddspace=.5em
	\renewcommand{\arraystretch}{1.6}

	\begin{center}
		\small
		\caption{Parameters used in the CosmoMC analysis and their prior ranges.
		\label{tab:priors}}\vspace{0.5cm}
		
		\begin{tabular}{lll}

			\toprule
		Parameter  & Prior & Physical Meaning \\
			\midrule
		  $\Omega_b h^2$  & $[0.005,0.1]$  & baryon density today \\[-1mm]
		   $\Omega_c h^2$  & $[0.001,0.99]$ & cold dark matter density today \\[-1mm]
		  $100 \hskip 1pt \theta_{\mathsmaller{\rm MC}}$  & $[0.5,10.0]$  &  angular sound horizon~\cite{Lewis:2002ah} \\[-1mm]
		  $\tau$  & $[0.01,0.8]$  & optical depth to reionization \\[1mm]
		\midrule
		 $\ln(10^{10}A_s)$  & $[2.7,4.0]$  &   scalar amplitude \\[-1mm]
		 $c_s$  & $[0.02,1.0]$& speed of sound \\[-1mm]
		 $\varepsilon_1$ & $[0.001,0.5]$ & Hubble flow parameter (HFP) \\[-1mm]
		 $\varepsilon_2$ & $[-0.5,0.5]$ & HFP \\[-1mm]
		 $\varepsilon_3$ & $[-0.1,0.1]$ & HFP \\[-1mm]
		 $\delta_1$ & $[-0.5,0.5]$ & sound flow parameter (SFP)\\[-1mm]
		 $\delta_2$ & $[-0.1,0.1]$ & SFP\\[1mm]
		 	\bottomrule
		\end{tabular}
	\end{center}
	\end{table}
 
\begin{figure}[htbp]
\begin{center}
\includegraphics[scale=0.41]{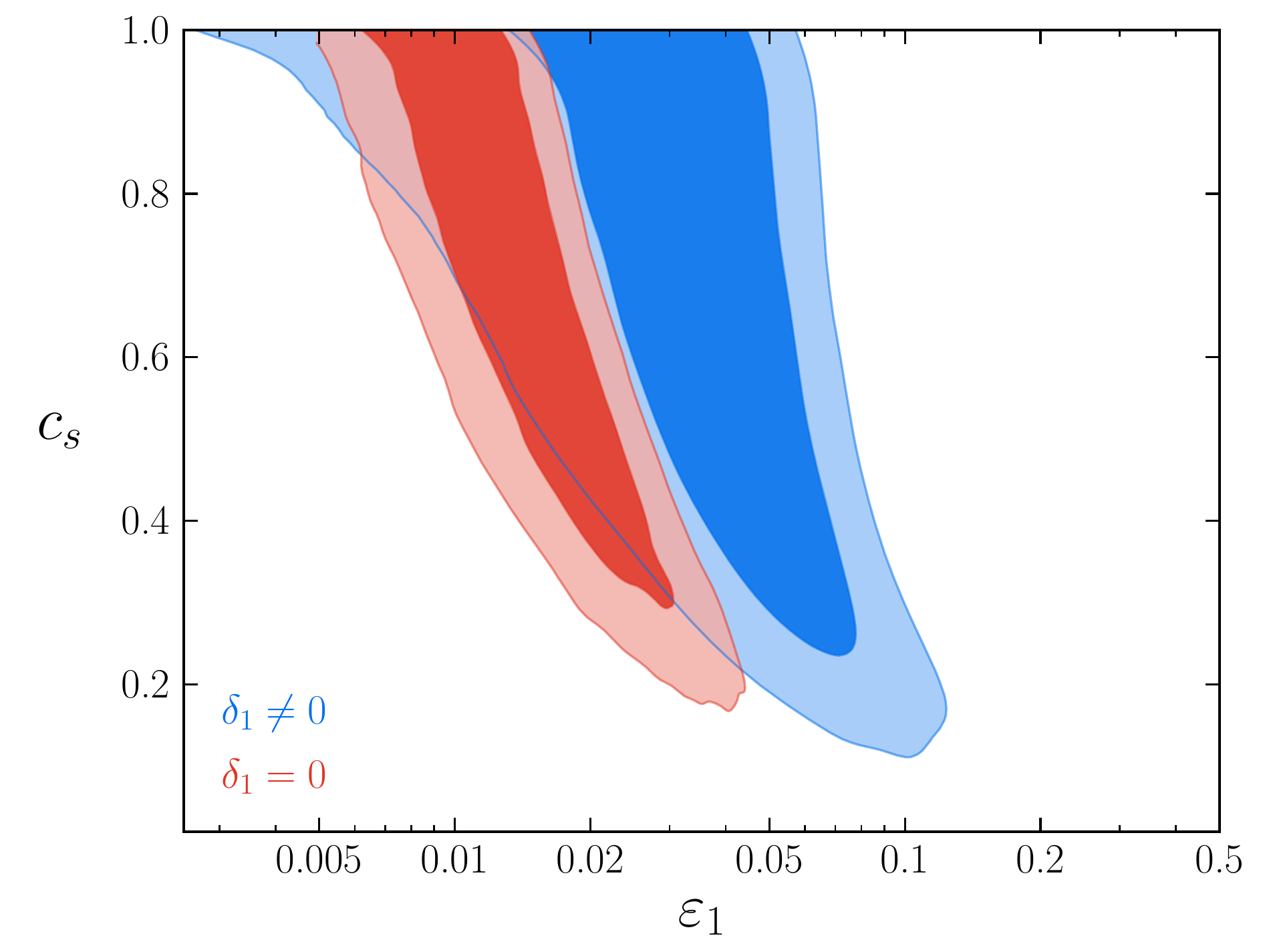}\hspace{0.4cm}
\includegraphics[scale=0.41]{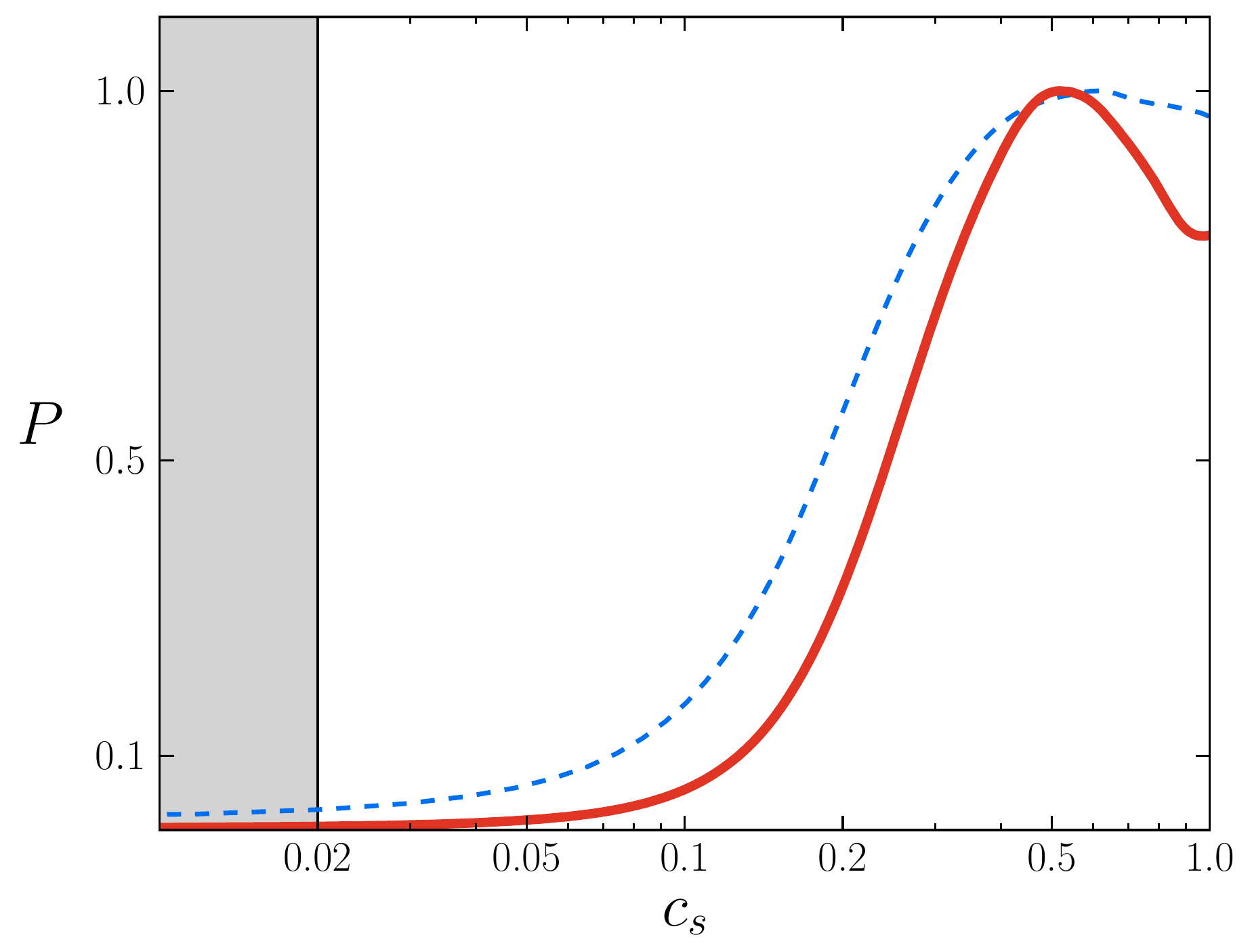}
\vskip -8pt
\caption{{\it Left:} 68\% and 95\% confidence contours of the marginalized posterior probability distribution for the parameters $c_s$ and $\varepsilon_1$.  The red shows shows that case with $\delta_1 = 0$, while the blue allows for $\delta_1 \ne 0$. {\it Right:} Marginalized posterior probability distribution for $c_s$, for $\delta_1=0$ (solid red line) and $\delta_1 \ne 0$ (dashed blue line).  The Planck exclusion, $c_s >0.02$, is shown in grey. }
\label{fig:run1}
\end{center}
\end{figure}

Our theoretical model consists of the standard 
 cosmological parameters of the $\Lambda$CDM model,
$\theta_{\rm lcdm} = \{ \Omega_bh^2, \Omega_c h^2, \theta_{\mathsmaller{\rm MC}}, \tau\}$, as well as the inflationary parameters $\theta_i = \{A_s,c_s, \varepsilon_1, \varepsilon_2,\varepsilon_3, \delta_1, \delta_2\}$ characterizing the initial conditions. We also consider various subsets of the inflationary parameters.
We modified the Boltzmann code {\tt CAMB}~\cite{Lewis:1999bs} to take the spectra of \S\ref{sec:2nd} as input and compute the joint likelihood $P(d|\theta)$, i.e.~the probability of the data $d$ given the model parameters $\theta$.
We use uniform prior probabilities $P(\theta)$ with prior ranges listed in table~\ref{tab:priors}.
The parameter space was sampled by a Markov-Chain Monte Carlo (MCMC) method using the publicly available software package {\tt CosmoMC}~\cite{Lewis:2002ah}. We ran twenty chains until the variation in the means of the chains was small relative to the standard deviation (using $R-1 < 0.02$ in the Gelman-Rubin~\cite{Gelman:1992zz} criterion).
In the plots below we present the posterior probabilities $P(\theta | d)$ for our model parameters.
For most of the analysis the cosmological parameters $\theta_{\rm lcdm}$ are fixed to their best-fit Planck values~\cite{Ade:2013zuv}, except in \S\ref{sec:rob} where we marginalize over them and include baryon acoustic oscillation (BAO) data~\cite{Cole:2005sx,Eisenstein:2005su} to break degeneracies.  Let us remind the reader that all bounds quoted here are 95\% confidence limits unless otherwise stated.
\vskip 4pt
The current best limit on $c_s$ comes from the absence of primordial non-Gaussanity in the Planck data: $c_s > 0.02$.  This limit assumes a constant sound speed.  To make a direct comparison\footnote{The Planck limit includes a marginalization over the parameter $M_3$ discussed in \S\ref{sec:stablehier}.  Since this parameter does not contribute to the power spectrum our bound does not need to be adjusted to make this comparison.} with the Planck analysis, we should therefore take the space of parameters to be $\theta_{i} = \{c_s, \varepsilon_1,\varepsilon_2 \}$, while setting $\varepsilon_3=\delta_{1,2} =0$.  We will consider this case first, and then discuss what happens when we add additional inflationary parameters.
The one-dimensional marginalized posterior probability distribution for $c_s$ is shown in fig.~\ref{fig:run1} and the inferred bound on $c_s$ is
\beq
 c_s > 0.25 \quad\quad {\rm for}\ \ \delta_1=0\ . \label{equ:csbound2a} 
\eeq
We remind the reader that this bound was derived from the BICEP2 likelihood {\it without} foreground subtraction. Including the effects of foregrounds would lower the bound, or might even remove it completely if no primordial tensor signal survives further scrutiny.  A complete analysis will have to await the improved likelihoods that will appear with the Planck/BICEP joint analysis.

While the split between $\varepsilon_{1,2}$ and $\varepsilon_{3}$ is well-motivated by the order at which they appear in the slow-roll expansion, setting $\delta_1 =0$ is a fairly artificial choice.  For example, in (\ref{equ:ns}) we see that $\delta_1$ contributes to the spectral index on the same footing as $\varepsilon_{1,2}$.  
As our {\it baseline} analysis, we will therefore take $\theta_{i} = \{c_s, \varepsilon_1,\varepsilon_2, \delta_1 \}$.   
The left panel of fig.~\ref{fig:run1} shows the marginalized posterior probability distribution for the parameters $c_s$ and $\varepsilon_1$ for both the baseline analysis and the case $\delta_1 = 0$.  We see that larger values of $\varepsilon_1$ are allowed if we marginalize over finite $\delta_1$.
This feature can be understood in terms of the two degeneracies discussed in \S\ref{sec:degeneracies}.  For $\varepsilon_1 > 0.01$, one of the degeneracies must be enacted to accommodate the observed value of $n_s$.  When $\delta_1=0$, the only option is $\varepsilon_2 \simeq -2 \varepsilon_1$, which is constrained by the limit on $\alpha_s$. On the other hand, if $\delta_1 \simeq -2\varepsilon_1$ the constraint from the running is satisfied provided $|\varepsilon_2| \ll \varepsilon_1$, in which case the bound on $\varepsilon_1$ is relaxed.
The one-dimensional posterior for $c_s$ is now given by the dashed curve in fig.~\ref{fig:run1}.  The bound on $c_s$ is
\beq
 c_s > 0.21 \quad\quad {\rm for}\ \ \delta_1 \ne 0\ . \label{equ:csbound2b} 
\eeq
We note that both (\ref{equ:csbound2a}) and (\ref{equ:csbound2b}) are
somewhat stronger than the corresponding analytic bounds in \S\ref{sec:degeneracies}.  The goal for the remainder of this section is to understand the reason for these stronger bounds and to determine their robustness to changes in our priors and variations of the data sets. As we shall see, much of the difference is driven by the anomalies in the low-$\ell$ data.

\subsection{Robustness of the Bound}\label{sec:rob}

\begin{figure}[htbp]
\begin{center}
\includegraphics[scale=0.41]{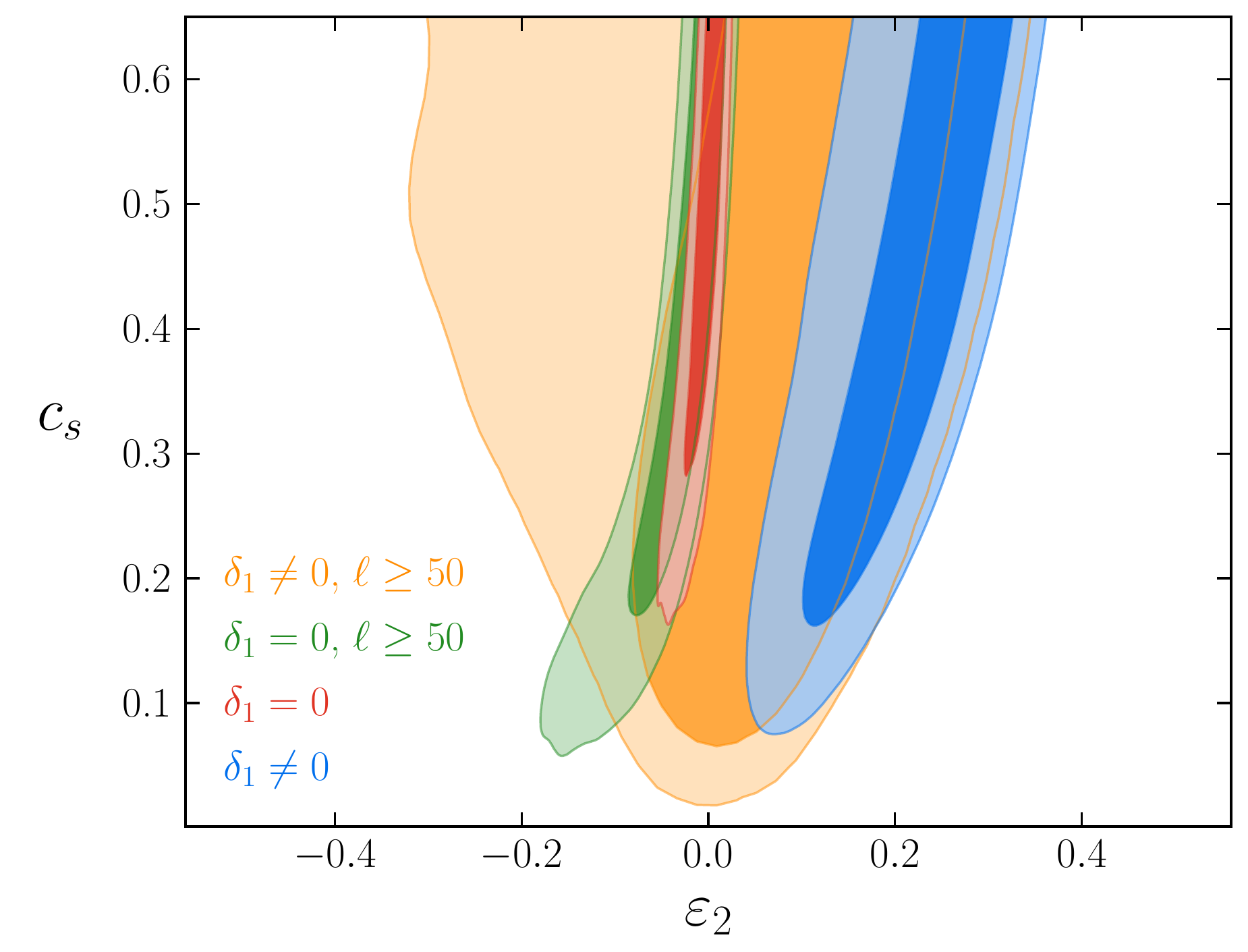}\hspace{0.4cm}
\includegraphics[scale=0.41]{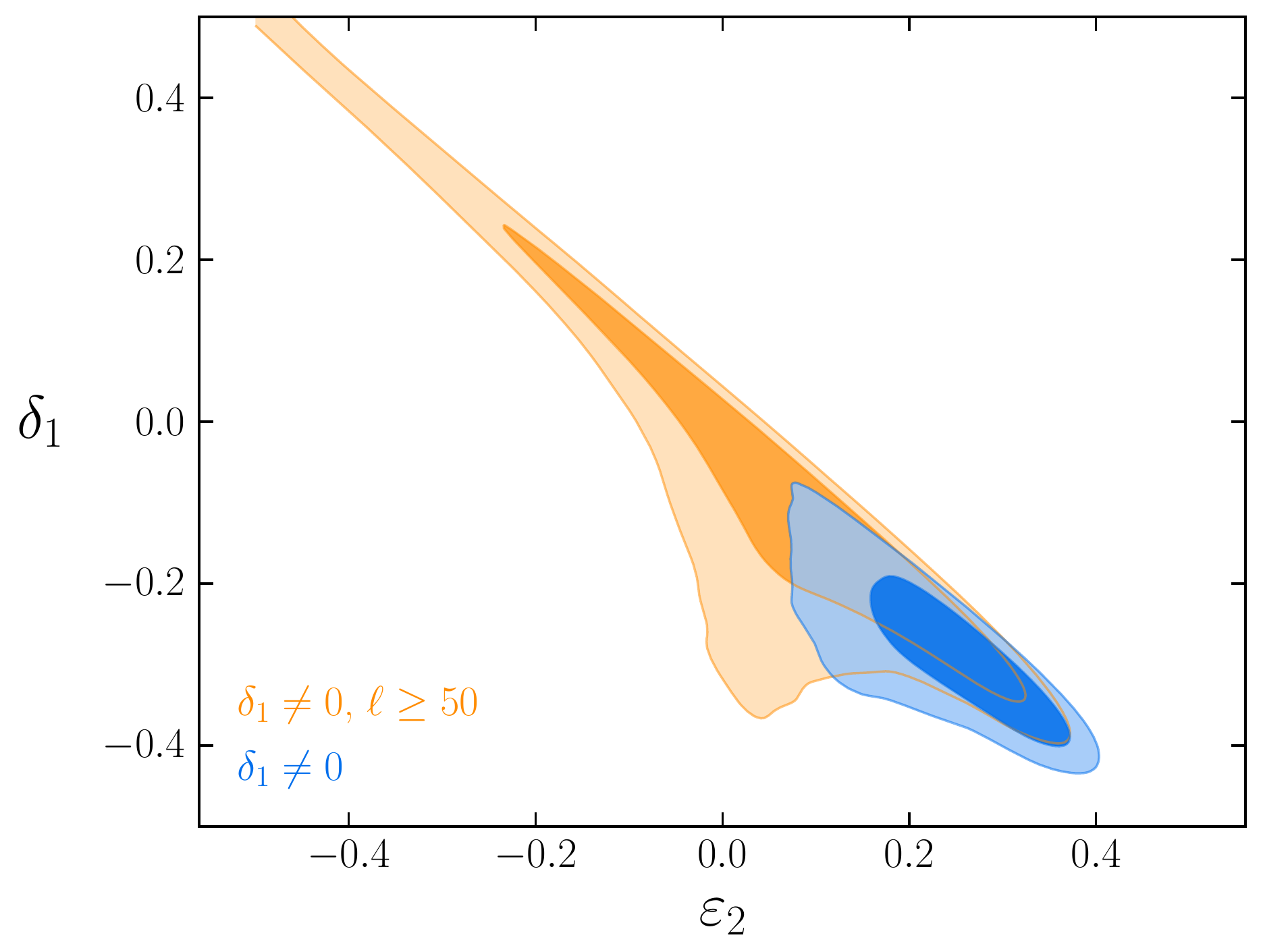}
\vskip -8pt
\caption{68\% and 95\% confidence contours of the marginalized posterior probability distributions. {\it Left:} contours for the parameters $c_s$ and $\varepsilon_2$ for $\delta_1= 0$ and $\delta_1 \neq 0$ both with and without the low-$\ell$ data. 
We note that the full data set disfavors large negative $\varepsilon_2$, while it becomes allowed when the low-$\ell$ data is excluded.  The contours for $\delta_1 \neq 0$ also show that the smaller values of $c_s$ push $\varepsilon_2 \to 0$, as we would expect from the constraint on $\alpha_s$ along the $\delta_1 \simeq -2\varepsilon_1$ degeneracy with large values of $\varepsilon_1$.  A similar feature does not appear when $\delta_1 = 0$, 
because in that case lower values of $c_s$ are allowed along the $\varepsilon_2 \simeq -2\varepsilon_1$ degeneracy.  
 {\it Right:} contours for the parameters $\delta_1$ and $\varepsilon_2$ with and without the low-$\ell$ data. The tendency for positive $\varepsilon_2$ disappears when the low-$\ell$ data is removed and large negative values of $\delta_1$ are allowed when $\varepsilon_2$ is small.}
\label{fig:degen1}
\end{center}
\end{figure}

As we described in \S\ref{sec:degeneracies}, small values of $c_s$ could be made consistent with a large tensor-to-scalar ratio, e.g.~$r > 0.13$, if $\varepsilon_2 \ll 0$.  However, this would require that the running is positive, $\alpha_s = -2\varepsilon_1 \varepsilon_2 > 0$, whereas it is well-known that negative running, $\alpha_s < 0$, improves the fit to the low-$\ell$ CMB temperature data~\cite{Ade:2013zuv}.  When $\delta_1 =0$, this tension then produces a much stronger constraint on $\varepsilon_2$, along the $\varepsilon_2 \simeq -2\varepsilon_1 <0$ degeneracy, than we would expect from the error on $\alpha_s$ alone (see the red contours in fig.~\ref{fig:degen1}).  This constraint is eliminated when we allow for $\delta_1 \neq 0$, and instead there is a clear tendency from the low-$\ell$ multipoles to prefer $\varepsilon_2 > 0$  (see the blue contours in fig.~\ref{fig:degen1}).  

\begin{figure}[h!]
\begin{center}
\includegraphics[scale=0.4]{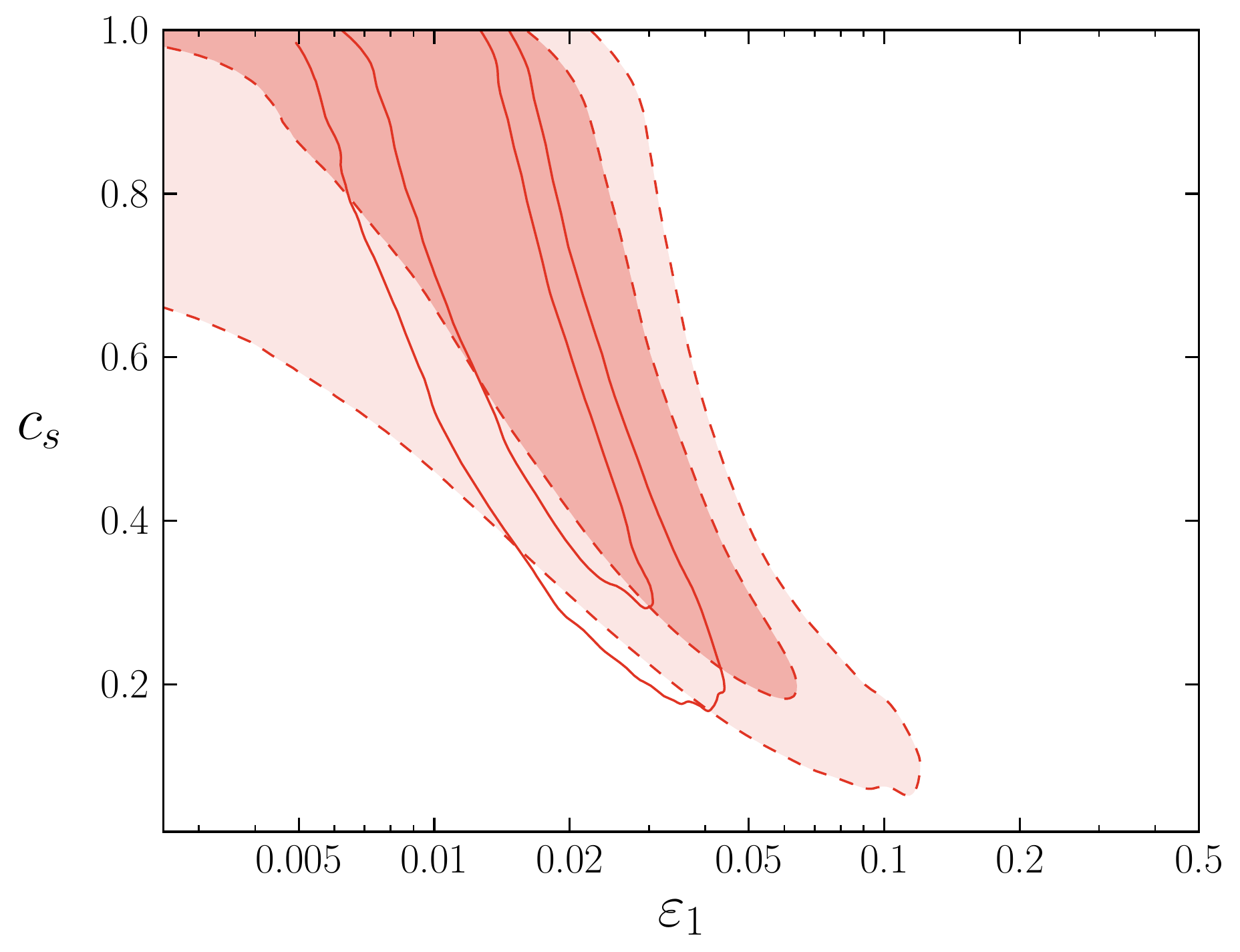}
\vskip -8pt
\caption{68\% and 95\% confidence contours for $\delta_1 =0$  with and without the low-$\ell$ data are shown in solid and dashed lines, respectively.  The contours without the low-$\ell$ data are filled in red.  We see that removing the low-$\ell$ data allows for larger values of $\varepsilon_1$, which favors smaller $c_s$.  This is consistent with the observation that a preference for negative $\alpha_s$ is responsible for the strong limits on $\varepsilon_1$ and $c_s$ shown in~fig.~\ref{fig:run1}.}
\label{figraf}
\end{center}
\end{figure}

To test how much our bounds depend on the anomalously low power in the low-$\ell$ CMB data, we repeat the analysis without the low-$\ell$ likelihoods.
Removing the low-$\ell$ data eliminates the preference for $\alpha_s < 0$ and hence allows $\varepsilon_2$ to take on a wider range of values.  
In particular, for $\delta_1=0$, excluding the low-$\ell$ multipoles allows for somewhat larger (and negative) values of $\varepsilon_2$ (and hence smaller values of $c_s$). This can be seen by comparing the green and red contours in fig.~\ref{fig:degen1} or by considering the range of allowed $\varepsilon_1$ with and without the low-$\ell$ data in fig.~\ref{figraf}.  As expected, the new bound on $c_s$ is weaker than in the case of the full data set,
\beq
 c_s > 0.14 \quad\quad  {\rm for}\ \ \delta_1 = 0,\ \ell \ge 50\ . \label{equ:csbound2c} 
\eeq
For $\delta_1 \ne 0$, removing the low-$\ell$ data has the effect of allowing large values of $\varepsilon_1$ which lower the bound on $c_s$, combined with small values of $\varepsilon_2$ to satisfy the constraint from the running (see the orange contours in the left panel of fig.~\ref{fig:degen1}).  
In this case the main degeneracy relevant to small values of $c_s$ is 
$\delta_1 \simeq -2\varepsilon_1$, where the combination of large negative $\delta_1$ and small $\varepsilon_2$ is allowed (see the orange contours in the right panel of fig.~\ref{fig:degen1}). 
As a result, the bound on $c_s$ is also weaker, 
\beq
 c_s > 0.15 \quad\quad  {\rm for}\ \ \delta_1 \ne 0,\ \ell \ge 50\ . \label{equ:csbound2d} 
\eeq
The expressions in (\ref{equ:csbound2c}) and (\ref{equ:csbound2d}) essentially match the corresponding analytic bounds in \S\ref{sec:degeneracies}. This strongly suggests that the previous difference between the analytic results and the full likelihood analysis was driven by the low-$\ell$ data.\footnote{Notice that allowing for $\delta_1 \neq 0$  in the analysis without the low-$\ell$ data does not produce a significant change in the bound on $c_s$. See also \S\ref{sec:degeneracies}.}

\begin{figure}[h!]
\begin{center}
\includegraphics[scale=0.45]{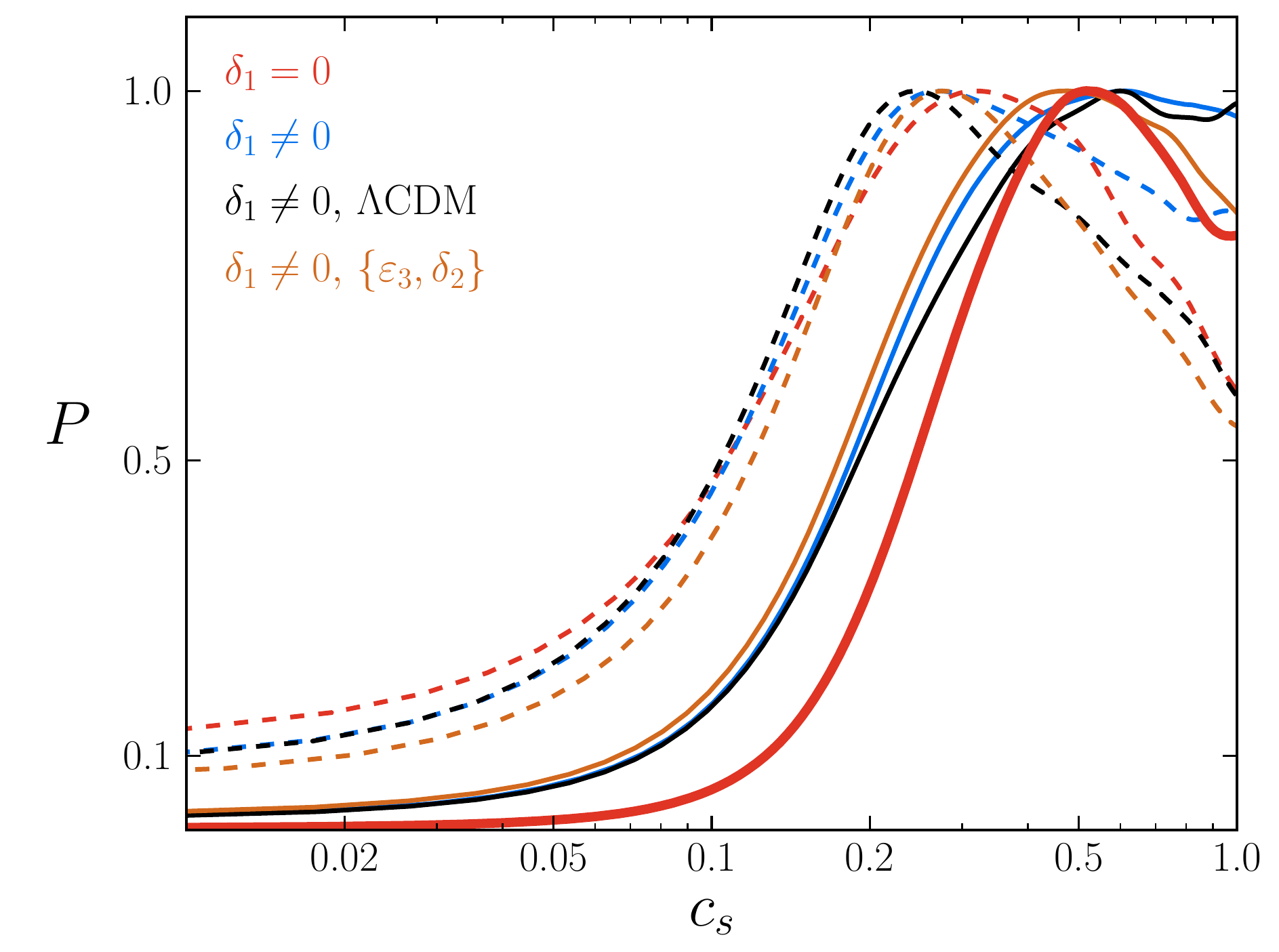}
\vskip -8pt
\caption{Marginalized posterior probability distributions for $c_s$ for the complete CMB data (solid lines) and without the low-$\ell$ data (dashed lines).  The labels `$\Lambda$CDM' and `$\{\varepsilon_3,\delta_2\}$' denote marginalizations over cosmological parameters and higher-order flow parameters, respectively.}
\label{fig:run1b}
\end{center}
\end{figure}

Finally, we wish to explore how robust our bounds are to changes in our priors.
First, we consider whether there is a significant difference between fixing the cosmological parameters of the $\Lambda$CDM model (as we did so far) and marginalizing over them.
The posteriors for $c_s$ for both cases are shown in fig.~\ref{fig:run1b} 
 and we find little change to our results.
Second, we allow for the higher-order flow parameters $\varepsilon_3$ and $\delta_2$ in the initial conditions. From (\ref{equ:alphas}) we see that both parameters could potentially cancel the contribution to running from $-2\varepsilon_1 \varepsilon_2$.  However, in order to achieve such a cancelation, $\varepsilon_3$ or $\delta_2$ would need to be large.  Allowing for such large values is inconsistent with a hierarchical structure of the slow-roll parameters.  Furthermore, third- and higher-order slow-roll corrections which we are neglecting are not necessarily small under such circumstances.  If we restrict the range of parameter to be $\{\delta_2, \varepsilon_3\} \in [-0.1,0.1]$, we find little effect on $c_s$; see fig.~\ref{fig:run1b} and table~\ref{tab:bounds}.  

 \begin{table}[h!]

	\heavyrulewidth=.08em
	\lightrulewidth=.05em
	\cmidrulewidth=.03em
	\belowrulesep=.65ex
	\belowbottomsep=0pt
	\aboverulesep=.4ex
	\abovetopsep=0pt
	\cmidrulesep=\doublerulesep
	\cmidrulekern=.5em
	\defaultaddspace=.5em
	\renewcommand{\arraystretch}{1.6}
 	
 	\begin{center}
		\small
		\caption{Bounds on the sound speed with the low-$\ell$ likelihoods ({\tt CMBAll}) and without (\xcancel{\tt LowLike}). 
		\label{tab:bounds}}\vspace{0.5cm}
		
		\begin{tabular}{l cc}

			\toprule
		 & {\tt CMBAll} & $\xcancel{\tt LowLike}$ \\
			\bottomrule
		  \ \ \ $\delta_1 =0$  & \ $c_s >0.25$\  & \ $c_s > 0.14$\ \ \ \\[-1mm]
		  \ \ \  $\delta_1 \ne 0$  & \ $c_s > 0.21 $\ & \ $c_s > 0.15$\ \ \  \\[-1mm]
		  \ \ \  $\delta_1 \ne 0, \Lambda{\rm CDM}$  & \ $c_s > 0.21$\  &  \ $c_s > 0.13$\ \ \   \\[-1mm]
		  \ \ \  $\delta_1 \ne 0, \{\varepsilon_3, \delta_2\}$  & \ $c_s > 0.20$\  & \ $c_s > 0.13$\ \ \  \\[1mm]
		 	\bottomrule
		\end{tabular}
	\end{center}
	\end{table}

\subsection{Prospects of Future Observations}

Our bound on $c_s$ is approaching the critical value, $(c_s)_\star = 0.47$, yet experimental improvements are required in order to reach it. The reason our bound is not stronger is due to the well-understood degeneracies of \S\ref{sec:degeneracies}. To improve the bound, we therefore have to search for observables that break these degeneracies. We will briefly comment on a few possibilities:

\begin{itemize}
\item A measurement of the tensor tilt, $n_t =-2 \varepsilon_1$, determines $\varepsilon_1$ directly, and therefore breaks all of the important degeneracies that limit the bound on $c_s$.  
 If $r\simeq 0.13$, then to get to $c_s > 0.47$ at the 95\% confidence level, would require $\sigma_{n_t} \sim 0.01$. Unfortunately, to achieve this sensitivity is essentially impossible with CMB measurements alone~\cite{Dodelson:2014exa}. Realistic projections for future constraints on $n_t$ may get down to $\sigma_{n_t} \sim 0.1$~\cite{Wu:2014hta}, which is not sufficient to significantly improve the bound on $c_s$.  

\item For a constant sound speed, $\delta_1=0$, the main degeneracy that limits the bound on $c_s$ is $\varepsilon_2 \simeq - 2 \varepsilon_1$.  Improving the constraint on $\alpha_s \simeq 4 \varepsilon_1^2$ then limits the size of $\varepsilon_1$, leading to a stronger bound on~$c_s$.  In particular, if $r \simeq 0.13$, we need $\varepsilon_1 < 0.017$ to get $c_s > 0.47$. To attain this level of sensitivity requires $\sigma_{\alpha_s} < 4\times 10^{-4}$. Notice that, for such an experiment, the precise values of $n_s$ and $r$ could in principle exclude $c_s =1$.  Bounding the running at the level $\sigma_{\alpha_s} \sim 10^{-3}$ is within reach of near-term CMB observations~\cite{Wu:2014hta} and future galaxy surveys~\cite{Takada:2005si, Adshead:2010mc}, although some improvement beyond these projections may be necessary to reach the critical value $(c_s)_\star =0.47$.

\item For a varying sound speed, $\delta_1 \ne 0$, any constraints on $n_s$ and $\alpha_s$ can be satisfied if
$\delta_1 \simeq - 2\varepsilon_1$, provided $\varepsilon_2$ remains small.  As a result, the bound on $c_s$ arises only from the analytic properties of $r(\varepsilon_1, \varepsilon_2, c_s)$ along the degeneracy.  However,  large values of $r$ are only consistent with small $c_s$ if $\delta_1 \simeq -2\varepsilon_1$ is relatively large.  For example, to achieve
 $r \simeq 0.13$ with $c_s \simeq 0.14$ requires $\varepsilon_1 \simeq 0.1$ and $\delta_1 \simeq -0.2$.  These values of $\delta_1$ are sufficiently large that they may be detectable.  Specifically, the value of $c_s$ relevant for the constraint from $r$ is set at $\ell \sim 10^2$ \cite{Ade:2014xna}, whereas most of the modes in a given data set are at $\ell \sim \ell_{\rm max} \gg 10^2$.  Using $\delta_1 = \dot c_s/(c_s H)$, we have
\beq
c_s(\ell_{\rm max}) = c_s(\ell=10^2)  \left(\frac{\ell_{\rm max}}{10^2}\right)^{\delta_1} = 0.07 \times \frac{c_s(\ell = 10^2)}{0.14}  \left(\frac{\ell_{\rm max}}{3000}\right)^{\delta_1(\ell = 10^2)/(-0.2)}\ . \label{equ:csScale}
\eeq
The value of $c_s =0.07$ at $\ell_{\rm max} \simeq 3000$, translates into $\fnl^{\rm equil} \simeq -52$, which may be reachable with future LSS surveys and/or high-resolution CMB satellites.\footnote{Projections for these experiments (e.g.~\cite{Sefusatti:2009xu}) give $\sigma(\fnl^{\rm equil}) \sim 10$.  Although these forecasts are for constant~$c_s$, most of the information in the surveys is at high $\ell$ (or $k$) and therefore insensitive to the variation of $c_s$. In particular, given that the number of modes scales like $\ell^2$ (or $k^3$), the scale dependence in (\ref{equ:csScale}) is much weaker than the scale dependence of the signal-to-noise.}  Similarly, reaching the threshold $c_s(\ell = 10^2) = 0.47$, under the same assumptions, requires sensitivity to $\fnl^{\rm equil} \simeq 5 (\fnl^{\rm equil})_\star = -4.4$.

\end{itemize}

\section{A Theoretical Threshold}
\label{sec:threshold}

Thresholds are an essential part of physics and targets for experimental searches. The LHC was built in part to explore the unitarity threshold derived from the Fermi scale, $\Lambda \simeq 4\pi (\sqrt{2}G_F)^{-1/2} \sim$ TeV, including the possibility (ultimately disfavored by the discovery of a weakly coupled Higgs boson) of strong dynamics, such as `technicolor' or `composite Higgs' models, playing a role in electroweak symmetry breaking. 
In cosmology, on the other hand, current observational evidence has not yet ruled out inflationary backgrounds driven by strongly coupled dynamics. In this section, we derive a (perturbative) unitarity threshold for the sound speed, $(c_s)_\star=0.47$, and the associated equilateral non-Gaussianity, $(\fnl^{\rm equil})_\star = -0.93$, within the framework of the EFT of inflation~\cite{Cheung:2007st}.
In Section~\ref{sec:NG}, we will review well-motivated scenarios that may produce a signal above the order-one threshold on equilateral non-Gaussianity.

\subsection{Sound Waves in the Early Universe}

The theory of the Goldstone boson $\pi(t,{\boldsymbol x})$ was introduced in \cite{Creminelli:2006xe, Cheung:2007st}. In the so-called {\it decoupling limit}, where the mixing with gravity is ignored, $\pi$ parameterizes the breaking of time translations in a quasi-de Sitter background with~a slowly evolving Hubble parameter $H(t)$.
The Goldstone boson also characterizes adiabatic density perturbations, which by definition, can be set to zero through the time diffeomorphism $t \to t-\pi(t,{\boldsymbol x})$. In this {\it unitarity gauge}, the fluctuations are described by the curvature perturbation $\zeta$, which at leading order is related to $\pi$ via $\zeta = - H\pi$. In this framework, standard slow-roll models 
are described by the Lagrangian: ${\cal L}^{(0)}_\pi = \Mp^2 \dot H(t) (\partial_\mu \pi)^2$. Different slow-roll models are distinguished only by the function $H(t)$. The first natural deformation of slow-roll inflation can then be parameterized as  \cite{Cheung:2007st}
\begin{align}
S_\pi &\,=\, \int \d^4 x\, a^3 \left\{ \Mp^2 \dot H(t)  (\partial_\mu \pi)^2  + 2M_2^4(t) \left[\dot \pi^2 + \dot\pi^3 - \frac{\dot \pi (\partial_i \pi)^2}{a^2}\right]  
 \right\}\ , \label{equ:S}
\end{align}
where we have only shown terms up to cubic order in $\pi$.
Turning on finite $M_2$ induces a sound speed for the Goldstone modes
\beq
c_s^2 \equiv \frac{\Mp^2 \dot H}{\Mp^2 \dot H - 2 M_2^4} \ .
\eeq 
Crucially, a small sound speed (large $M_2$) simultaneously affects different orders in the Goldstone action.
In particular, the contributions to the quadratic, $\dot \pi^2$, and cubic, $\dot \pi (\partial_i \pi)^2$, Lagrangian are not independent, but are related by a non-linearly realized symmetry~\cite{Cheung:2007st}.\footnote{The size of $\dot \pi^3$ is not fixed by the value of $c_s$, and can be adjusted by another term in the EFT; see \S\ref{sec:stablehier}.} This feature allows power spectrum measurements to constrain the strength of certain interactions in the theory.

\subsection{Energy Scales}

Two of the important energy/frequency scales in the theory of the Goldstone bosons are the {\it freeze-out scale} ($H$), and the {\it symmetry breaking scale}\footnote{We use $f_\pi$ to denote the symmetry breaking scale to highlight the analogy with the pion decay constant in chiral perturbation theory.} ($f_\pi$). The freeze-out of the Goldstone dynamics happens universally when the physical frequency of a mode ($\omega$) drops below the Hubble scale.\footnote{Here, we assume that the parameters that control the dynamics of the Goldstone boson vary slowly with time or essentially remain constant. When the time variation becomes important, modes do not necessarily freeze-out at $\omega \simeq H$. Nevertheless, the following discussion on the role of the relevant energy scales remains valid~\cite{Behbahani:2011it, Flauger:2013hra, Adshead:2014sga,Cannone:2014qna}.}
(The corresponding momentum depends on the dispersion relation.) The symmetry breaking scale, on the other hand, is model-dependent. For models with a sound speed, one finds~\cite{Baumann:2011su}
\beq
f_\pi^4 \equiv 2 \Mp^2 |\dot H| c_s\ . \label{equ:fpi}
\eeq
In the slow-roll limit, $c_s \to 1$, this definition reduces to $f_\pi^2 \to \dot \phi$. At energies below $f_\pi$ a description in terms of the Goldstone boson is appropriate.
Assuming vacuum fluctuations, the ratio of $H$ and $f_\pi$ is fixed by the amplitude of curvature perturbations~\cite{Baumann:2011ws}
\beq
\label{eq:dz}
\Delta_\zeta^2 =  \frac{1}{4\pi^2} \left(\frac{H}{f_\pi}\right)^{2\delta_\pi + 2}  \ ,
\eeq
where $\delta_\pi$ is the scaling dimension of $\pi$.  For sound speed models, we have $\delta_\pi = 1$ and the amplitude of curvature perturbations is enhanced by an inverse power of the sound speed, $\Delta_\zeta^2 \propto c_s^{-1}$. The observed value for \eqref{eq:dz}, $\Delta_\zeta^2 = 2.2 \times 10^{-9}$, determines 
\beq 
f_\pi \simeq 60\hskip 1pt H\ . 
\eeq

  \begin{figure}[htbp]
\begin{center}
\includegraphics[width=0.5\textwidth]{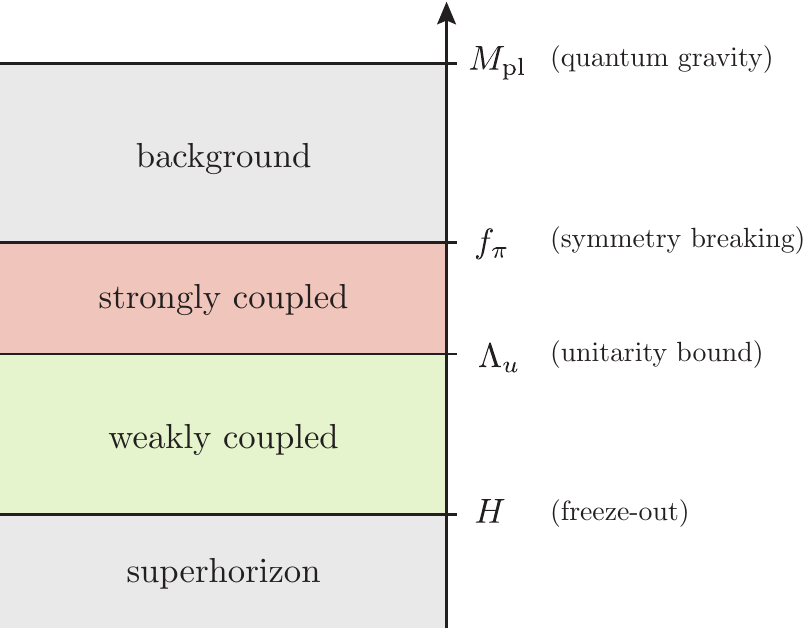}
\caption{Graphical illustration of the relevant energy scales in the EFT of inflation. Whether $\Lambda_u$ is above or below $f_\pi$ is an important qualitative distinction.}
\label{EnergyScales}
\end{center}
\end{figure}

Armed with a theory of the perturbations, a natural question is at what scale this theory becomes strongly coupled.\footnote{If we assume that a fundamental scalar field $\phi$ produces the inflationary background, then a sound speed for the fluctuations arises from higher-derivative corrections to the kinetic term~\cite{Creminelli:2003iq}
\beq
{\cal L} = -\frac{1}{2}(\partial \phi)^2 + \frac{(\partial \phi)^4}{M^4} + \cdots\ . \label{equ:L}
\eeq
For $\dot \phi < M^2$ this is a perturbative correction to the slow-roll dynamics, and as such $c_s \simeq 1$. 
On the other hand, for $\dot \phi > M^2$, the higher-order terms that are hidden in the ellipses in (\ref{equ:L}) become relevant and the background is non-perturbative. 
The computation of $\fnl^{\rm equil}$ is not reliable from the term displayed in \eqref{equ:L} alone, and the relevant scale for the validity of the expansion must be sought within the theory of the perturbations. At the same time, the theory of the Goldstone boson allows for more complicated situations where fluctuations are not necessarily associated with a fundamental scalar field.} 
To answer this one re-writes \eqref{equ:S} as \cite{Cheung:2007st,Baumann:2011su} 
\beq
\label{calL}
S_\pi = \int \d t \hskip 2pt \d^3 \tilde x\, a^3 \left\{ -\frac{1}{2} (\tilde\partial_\mu \pi_c)^2 - \frac{1}{2\Lambda^2}\left[  \frac{\dot\pi_c(\tilde\partial_i\pi_c)^2}{a^2} - c_s^2 \dot \pi_c^3 \right]  \right\} \ . 
\eeq
Here, we have rescaled the spatial coordinates, $\tilde x^i = c_s^{-1} x^i$, to reinstall (fake) Lorentz invariance in the quadratic part of the action, defined $\pi_c \equiv f_\pi^2\hskip 1pt \pi$, and introduced the scale
\beq
\Lambda^4 \equiv  2 M^2_{\rm pl} |\dot H| \frac{c_s^5}{(1-c_s^2)^2}\ . \label{equ:LL}
\eeq
We see that the contribution of the operator $\dot \pi^3_c$ is suppressed for $c_s \ll 1$ and the dominant non-linearity comes from $\dot \pi_c (\tilde \partial_i \pi_c)^2$. For this reason, we will, for now, restrict to the effects of~$\dot \pi_c (\tilde \partial_i \pi_c)^2$. We will return to a discussion of the operator $\dot \pi_c^3$ in \S\ref{sec:stablehier}.

The high degree of Gaussianity of the observed CMB anisotropies implies that the Goldstone modes are weakly coupled at freeze-out, $\omega \simeq H$.  This requires that the scale $\Lambda$ is above the Hubble scale, so that higher-dimension operators are suppressed by powers of $H/\Lambda < 1$ at horizon crossing.
However, extrapolating to higher frequencies 
the effects of higher-order operators becomes more relevant, until the perturbative description breaks down at the {\it strong coupling scale} ($\Lambda$). Perturbative unitarity is violated at the nearby {\it unitarity scale}:\footnote{This scale is obtained from imposing partial wave unitarity in the quartic interaction: $\frac{1}{8(1-c_s^2)} \frac{1}{ \Lambda^4} \frac{(\tilde\partial_i\pi_c)^4}{a^4}  \subset {\cal L}^{(4)}_\pi\ .$
The computation may be found in appendix E of \cite{Baumann:2011su} after fixing a minor discrepancy with the numerical coefficient in eq.~(E.7), which should read: \beq 
a_0 = \frac{5}{3} \times \frac{1}{16\pi} \frac{(1-c_s^2)\omega^4}{2M^2_{\rm pl} |\dot H| c_s^5} =  \frac{5}{48\pi}\frac{\omega^4}{(1-c_s^2)\Lambda^4}\ .
\eeq
The condition $\omega^4 < \Lambda_u^4$ follows from $a_0 < 1/2$, as required by unitarity.}
\beq 
\Lambda_u^4 \equiv  \frac{24\pi}{5} (1-c_s^2)\hskip 1pt \Lambda^4\ . \label{equ:Lu}
\eeq 
Whether this happens above or below the symmetry breaking scale is an important qualitative distinction. 
 The theory is close to a weakly coupled slow-roll background if $\Lambda_u > f_\pi$, while for $\Lambda_u < f_\pi$ the Goldstone action involves a completion below the symmetry breaking scale, plausibly\footnote{Weakly coupled completions of models with $\Lambda_u < f_\pi$ were studied in \cite{Baumann:2011su,Tolley:2009fg,Cremonini:2010ua,Achucarro:2010da,Cespedes:2012hu,Burgess:2012dz}.  These models always involve new physics parametrically below the scale $\Lambda_u$ and may be observationally distinguishable from their strongly coupled counterparts.} signaling strongly coupled dynamics.

\subsection{A Critical Sound Speed}
\label{ssec:threshold}

The critical value $\Lambda_u = f_\pi$ is an interesting target for future experiments. Using (\ref{equ:Lu}) and (\ref{equ:fpi}), this threshold is related to a critical sound speed
\beq
\label{eq:th3}
\frac{24\pi}{5} \left(\frac{2 M_{\rm pl}^2 |\dot H| (c_s)_\star^5}{1-(c_s)_\star^2}\right) = 2 M_{\rm pl}^2 |\dot H| (c_s)_\star  \qquad \Rightarrow \qquad  (c_s)_\star = 0.47\ .
 \eeq
The threshold value $(c_s)_\star$ is still far from the Planck-only bound, $c_s >0.02$.\footnote{The bound on $c_s$ derived from the interaction $\dot \pi (\partial_i \pi)^2$ alone is $c_s > 0.04$~\cite{PlanckNG}.  The limit $c_s > 0.02$ includes a marginalization over an additional parameter $M_3$ (see \S\ref{sec:stablehier} for the definition of $M_3$). Since $M_3$ does not contribute to the two-point function, the limits presented in this paper naturally include such a marginalization and should therefore be compared to the weaker limit.} On the other hand, our analysis in this paper shows that any detection of primordial B-modes with $r > 0.01$ improvesthis constraint on $c_s$. Moreover, the bound we obtained using the BICEP2 data (without foreground subtraction), $c_s \geq 0.25$, is already remarkably close to the unitarity threshold.
\vskip 4pt
To relate (\ref{eq:th3}) to a threshold for the non-Gaussianity associated with the interaction $\dot \pi (\partial_i \pi)^2$ in (\ref{equ:S}), we use
\begin{align}
(\fnl^{\dot\pi(\partial_i\pi)^2})_\star & \,=\,  - \frac{85}{324}  \frac{1-(c_s)_\star^2}{(c_s)_\star^2}  \simeq - 0.93  \ . \label{EFT1tar} 
\end{align}
This experimental benchmark is still two orders of magnitude below the limit from the Planck bispectrum measurement~\cite{PlanckNG}, $\fnl^{{\dot\pi(\partial_i\pi)^2}} = 8 \pm 146$.  In the next section, we will describe the different types of physics involved in non-Gaussianity at or above the threshold given by \eqref{EFT1tar}.

\section{Physics above Threshold}
\label{sec:NG}

It is apparent that there is a large window between the current bounds on non-Gaussianity and the threshold, $|\fnl^{\rm equil}|_\star \simeq {\cal O}(1)$. 
This offers a wonderful opportunity to explore non-canonical models through measurements of non-Gaussianity.\footnote{One may worry that a threshold at $|\fnl^{\rm equil}|_\star \simeq {\cal O}(1)$ could be experimentally elusive. Indeed, future CMB observations will not be able to reach this threshold~\cite{Baumann:2008aq}. On the other hand, the large number of modes that LSS observations in principle offer may in the future allow us to access smaller values of $\fnl^{\rm equil}$.  This will require a detailed understanding of secondary non-linearities in structure formation~\cite{Carrasco:2012cv, Carrasco:2013mua, Porto:2013qua, Angulo:2014tfa, Baldauf:2014qfa}.}
In this section, we survey a range of well-motivated theories that can produce a signal {\it at} or {\it above} the threshold.
(See also \cite{Creminelli:2014oaa, MatiasTalk, DAmico:2014cya,LeoTalk,DanTalk,RafaelTalk}, where combinations of the following scenarios have been reviewed.)

\subsection{One-Scale Models \& Strong Coupling} 

The standard example of strong coupling realized in nature is QCD. 
At low energies, chiral symmetry is broken and the dynamics of QCD is described by the associated (pseudo-)Goldstone bosons---the pions---whose interactions are described by the chiral Lagrangian.  The order parameter of the symmetry breaking is the vacuum condensate $\langle q\bar q\rangle \sim f_\pi^3$, with $f_\pi$ playing the role of the symmetry breaking scale. Higher-order interactions of the pions are also controlled by $f_\pi$, making chiral perturbation theory a `one-scale' theory. (For example, the chiral Lagrangian includes the quartic term $(12\pi^2f_\pi^4)^{-1} (\partial_\mu \pi)^4$. Cubic interactions are forbidden by parity.)
 
In cosmology, it may still be the case that interactions in the theory of the Goldstone boson in the EFT of inflation are controlled by a single scale, like for the pion(s) of QCD (without parity). Since inflation breaks time translations the analogy with chiral symmetry breaking would correspond to a (time-dependent) vacuum condensate with $\partial_t \langle q\bar q\rangle \sim~f_\pi^4$. In such a scenario the cubic interaction $ \dot\pi_c (\partial_i\pi_c)^2/f_\pi^2$ (up to an order-one numerical coefficient) would produce an order-one level of equilateral non-Gaussianity. This view is permitted by the data which still allows for a sound speed near the critical value, corresponding to one-scale models with $\Lambda_u \sim f_\pi$. The threshold  $(c_s)_\star=0.47$ is therefore a milestone for strongly coupled inflationary dynamics controlled by a single  scale.\footnote{One may be tempted to take the position that $0.47 \simeq 1$, and no further scrutiny is necessary. However, this attitude ignores well-known examples where the discrepancy between a weak coupling computation and its strong coupling counterpart is a small factor. For example, a computation of the entropy density $s$ in the quark-gluon plasma in terms of an effective `quasi-particle' description fits lattice data qualitatively well, i.e. $s\simeq 0.85 s_0$, with $s_0$ the entropy density of a non-interacting plasma, see e.g. \cite{Adams:2012th}. This small difference, $ (s_0-s)/s_0 \simeq 0.15$, has been taken as evidence of weak coupling at the relevant temperatures. However, a computation in a somewhat related theory (${\cal N}=4$ SYM) revealed that $s = f(\lambda) s_0$, with $f(\lambda) \to 3/4$ in the strong coupling limit $\lambda \gg 1$ \cite{Gubser:1998nz}. The proximity to the lattice computation suggests the quark-gluon plasma may be strongly coupled. In this example the function $f(\lambda)$ changes merely by a factor of $1/4$ when one extrapolates between weak and strong coupling.}  As we have shown in this paper, a confirmed detection of primordial tensor modes by the BICEP2 collaboration and improved bounds on the scalar spectrum by future experiments have the potential to probe these theories.

\subsection{Stable Hierarchies}\label{sec:stablehier}

Going beyond sound speed models, the action for the Goldstone boson in the EFT of inflation includes other cubic interactions~\cite{Cheung:2007st}
\begin{align}
S_\pi &\,=\, \int \d^4 x\, a^3 \left\{ - \Mp^2 \dot H \left[ \dot \pi^2 - \frac{(\partial_i \pi)^2}{a^2}\right] + 2M_2^4 \left[\dot \pi^2 - \frac{\dot \pi (\partial_i \pi)^2}{a^2} \right] + \left(2M_2^4 - \frac{4}{3} M_3^4\right) \dot \pi^3 \right. \nonumber \\[8pt]
& \left. \hspace{3cm} -\, \frac{H \bar M_1^3}{2}\frac{(\partial_i \pi)^2}{a^2} - \frac{\bar M_2^2 + \bar M_3^2}{2} \frac{(\partial_i^2 \pi)^2}{a^4} + \frac{\bar M_1^4}{2} \frac{\partial_i^2 \pi (\partial_j \pi)^2}{a^4} + \cdots \right\} \ . \label{equ:Spi}
\end{align}
As it has been pointed out in the literature, a small sound speed generates a large radiative correction to the parameter $M_3$, namely $M^4_3 \sim M^4_2/c_s^2$ \cite{Senatore:2009gt,Senatore:2010jy}. However, the converse is not the case, and various terms in \eqref{equ:Spi} may be large without inducing a significant modification of the sound speed. We briefly enumerate these possibilities: 

\begin{itemize}
\item $\boldsymbol{M_3}$  

It is technically natural to have large $M_3$ and small $M_2$~\cite{Baumann:2011nk, Creminelli:2014oaa, DAmico:2014cya}. To see this, let us set $M_2= \bar M_n = 0$ and only turn on finite $M_3$:
\beq
{\cal L}_\pi = M_{\rm pl}^2 \dot H (\partial_\mu \pi)^2  - \frac{4}{3}M_3^4\left(\dot \pi^3 - \frac{3}{2} \dot \pi^2 (\partial_\mu \pi)^2 + \cdots  \right)\ .
\eeq
We should be concerned that
a non-zero value of $M_2$ will be generated through loops
\beq
\includegraphicsbox[scale=0.8]{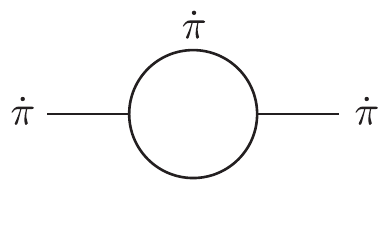}  \ \ +\ \ \includegraphicsbox[scale=0.8]{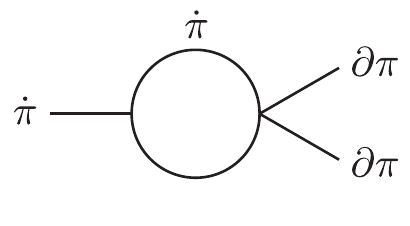} 
\ \ + \ \ \cdots \ \ =\ \ 2 M_2^4 \Big(\dot \pi^2 - \dot \pi (\partial_\mu \pi)^2  + \cdots \Big)\ . \eeq
However, running the loops all the way to the cutoff
$\Lambda_{\mathsmaller{\rm UV}} = (\Mp^2 |\dot H|)^{3/4}/M_3^2$, we only get $M_2^4 \sim \Mp^2 \dot H$, even if $M_3^4 \gg \Mp^2 |\dot H|$. A large value of $M_3$ is therefore consistent with having only small deviations from $c_s = 1$. This allows for large equilateral non-Gaussianity that is not constrained by measurements of the power spectrum. 

\item $\boldsymbol{\bar M_n}$ 

We can avoid the constraints on $c_s$ by allowing for large values of $\bar M_2$ and $\bar M_3$, which changes the dispersion relation to be that of ghost inflation~\cite{ArkaniHamed:2003uz}, namely $\omega = k^2/\rho$.  In this case, the relationship between $r$ and $\varepsilon_1$ is broken, and hence the tilt of the power spectrum does not constrain $c_s$.  The phenomenology of this setup was studied in detail by \cite{Senatore:2009gt}, which we briefly summarize below. 

The dispersion relation in this model takes the form:
\beq
\omega^2 = c_s^2 k^2 + \frac{k^4}{\rho^2}\ , \qquad {\rm where} \quad \quad \rho^2 \equiv \frac{2 \Mp^2 \dot H}{c_s^2  (\bar M_2^2+\bar M_3^2)} \xrightarrow{\ c_s \ll 1\ } \frac{4 M_2^4}{\bar M_2^2 +\bar M_3^2}\ .
\eeq
The $c_s^2$-term is negligible at horizon crossing provided that $c_s^2 \ll H/\rho$. Notice that this constraint is not trivially satisfied, because we must require that the modes propagate subluminally at horizon crossing, which implies $2 \left(H/\rho\right)^{1/2} < 1$. (The group velocity, given by $c_g \equiv {d \omega/d k} = {2 k/\rho}$, must be bounded: $c_g \le 1$.)

The non-linear dispersion enhances operators with a large number of spatial derivatives, and one can determine that \cite{Senatore:2009gt}
\begin{align}
\fnl^{\dot \pi (\partial_i \pi)^2} &= 0.25\, \frac{\rho}{H} \ ,\\[4pt]
\fnl^{\partial^2_i \pi(\partial_j \pi)^2} &= 0.13\,\frac{\bar M_1^3}{H(\bar M_2^2+ \bar M_3^2)}\ ,
\end{align}
for the two leading terms in the cubic action~\eqref{equ:Spi}.
It is interesting to note that self-consistency ($c_g < 1$) implies that 
\beq
\fnl^{\dot \pi (\partial_i \pi)^2}  > 1 \ .
\eeq
In other words, for this mechanism to be in operation we necessarily require non-Gaussianity above the threshold.

\end{itemize}

\subsection{Additional Degrees of Freedom}

The EFT of inflation has been extended to include couplings of $\pi$ to multiple (Goldstone-like) light fields~\cite{Senatore:2010wk}. One of the main signatures of these type of models is non-Gaussianity of the local type. The local shape, however, is highly constrained by the Planck data, $\fnl^{\rm loc} = 2.7 \pm 5.8$~\cite{PlanckNG}.  There is nonetheless a class of multi-field models, namely {\it single-clock models}, in which surfaces of constant density are effectively controlled by a single degree of freedom such that the consistency condition (implying a vanishing squeezed limit) is satisfied, thus bypassing the Planck constraint on local non-Gaussianity~\cite{LopezNacir:2012rm}. On the other hand, equilateral non-Gaussianity may still be generated even when $c_s=1$. We briefly describe two examples:

\begin{itemize}
\item {\it Dissipative dynamics.}---Various mechanisms have been proposed~\cite{Berera:1995ie, Green:2009ds} to produce the observed density perturbations through non-vacuum fluctuations in models with dissipative dynamics. Building upon ideas originally developed in the study of black hole absorption~\cite{Goldberger:2005cd, Porto:2007qi}, an EFT approach to generic classes of dissipative models was put forward in~\cite{LopezNacir:2011kk,LopezNacir:2012rm}. The main idea is to systematically couple the Goldstone boson to a dissipative sector described by composite operators ${\cal O}$, whose correlation functions are constrained by symmetries.\footnote{A similar approach was used to study dissipation in fluids~\cite{Endlich:2012vt}.}  These couplings can induce an effective friction term, $\gamma\dot\pi$, in the equations of motion.
For strongly dissipative systems, with $\gamma \gg H$, 
the power spectrum then depends on new parameters and is dominated by non-vacuum fluctuations.

Non-Gaussianity in dissipative models was studied in \cite{LopezNacir:2011kk,LopezNacir:2012rm}, where it was shown that the non-linearly realized symmetries require the presence of a non-linear term in the dynamics correlated with the leading order dissipation, $\gamma \dot \pi \to -\frac{1}{2} \gamma (\partial_i\pi)^2$, 
among other contributions. This term induces equilateral non-Gaussianity even when $c_s\simeq 1$, i.e. $\fnl^{\rm equil}\simeq -\gamma/4H$~\cite{LopezNacir:2011kk}. The bound from Planck on the equilateral shape then translates (roughly) into $\gamma \lesssim 10^2 H$. In addition, the bispectrum has another peak at the `folded' configuration: $k_1=k_2= \frac{1}{2}k_3$. This contribution is correlated with the size of the equilateral shape. For $\gamma \simeq {\cal O}(10)H$, one finds 
\beq 
\fnl^{\rm fold} \simeq  - \frac{1}{2} \fnl^{\rm equil}\ ,
\eeq 
with a negligible squeezed limit (dissipation erases memory quickly) \cite{LopezNacir:2012rm}. These features can be considered a smoking-gun for dissipative models, and a probe of the quantum nature of the primordial fluctuations \cite{RafaelTalk}.\footnote{Another way to obtain non-trivial squeezed limits is to assume a state other than the vacuum as an initial condition, i.e.~an {\it excited} state. The most popular choice involves Bogoliubov states. This possibility, however, is severely constrained by the data~\cite{Flauger:2013hra}. In dissipative models, on the other hand, the excited (semi-classical) state is reached dynamically.} 

\item {\it Quasi-single-field inflation.}---Equilateral non-Gaussianity also arises when the Goldstone mode couples to extra scalar fields with masses of order the Hubble scale, as in models of `quasi-single-field inflation'~\cite{Chen:2009zp}. Since massive fields decay outside the horizon, their interactions are localized at horizon crossing. This effect suppresses the interactions of modes with different wavelengths and produces an approximately equilateral shape for the bispectrum.  The squeezed limit of the bispectrum and the collapsed limit of the trispectrum depend on the mass of the field, allowing it to be distinguished from the equilateral shape produced by higher-derivative interactions in single-field inflation, or by non-adiabatic evolution as described above. Models of quasi-single-field inflation arise naturally in inflationary theories with spontaneously broken supersymmetry~\cite{Baumann:2011nk, Assassi:2013gxa, Craig:2014rta}.  

Quasi-single-field models can be generalized to include couplings to arbitrary additional sectors parameterized by composite operators ${\cal O}$, as in the case of dissipative dynamics~\cite{LopezNacir:2011kk}. In~\cite{Green:2013rd} the additional sector was taken to be a conformal field theory.  For this special case, the predictions are qualitatively similar to those of quasi-single-field inflation.
The phenomenology of more general cases is an open problem, but the expectation is that equilateral non-Gaussianity will be common.

\end{itemize}

\section{Conclusions and Outlook}

\label{sec:conclusions}

Inflationary models are characterized by a small number of important energy scales:
the freeze-out scale, $H$, the scale of time-variation of the background, $f_\pi$, and the scale(s) characterizing scalar self-interactions, $\Lambda$.   Cosmological observables in the scalar sector are only sensitive to the ratios $f_\pi/H$ and $\Lambda/H$.
 For example, the amplitude of curvature perturbations fixes $f_\pi/H$, while non-Gaussianity constrains $\Lambda/H$.  Measuring tensors, on the other hand, determines $H/\Mp$ and hence normalizes all energy scales relative to the Planck scale.  Famously, a large value of $f_\pi$ in Planck units, then correlates with a super-Planckian field excursion~\cite{Lyth:1996im,Baumann:2011ws} and an enhanced UV sensitivity of the inflationary model~\cite{Baumann:2008aq,Baumann:2014nda}. 
 Whether $\Lambda$ is above or below $f_\pi$ is an important qualitative distinction. To decide this question, in general, requires precision measurements of the non-Gaussianity corresponding to the interactions associated with the scale $\Lambda$. However,
in models with a sound speed $c_s$, the leading cubic interaction is related to the quadratic action by a non-linearly realized symmetry, and both $f_\pi$ and $\Lambda$ depend on $c_s$.  In this special case, power spectrum measurements can inform us about the interacting theory by constraining the value of~$c_s$.
  
\vskip 4pt
In this paper we have shown that consistency between an observable tensor amplitude and the near scale-invariance of the scalar spectrum enforces a lower bound on the sound speed. 
We found, analytically, that any observable tensor signal ($r>0.01$) constrains the sound speed above the bound obtained from the Planck temperature data alone. A joint analysis of the Planck temperature data and the BICEP2 B-mode measurement (without foreground subtraction) also supports another related conclusion: A future detection of primordial tensor modes with $r \gtrsim 0.1$ would lead to a bound on $c_s$ which is an order of magnitude stronger than current bounds derived from non-Gaussianity.
In such scenario, the resulting bound on $c_s$ would be tantalizingly close to a natural threshold,
\beq 
(c_s)_\star = 0.47 \ ,\label{equ:csStar}
\eeq 
namely the value of the sound speed for which the unitarity scale associated with the leading cubic interaction coincides with $f_\pi$ (see also~\cite{LeoTalk, DanTalk}). This corresponds to a related threshold of equilateral non-Gaussianity with amplitude 
\beq |\fnl^{\dot \pi (\partial_i \pi)^2}|_\star = 0.93 \ .\eeq 

Values of $c_s$ below (\ref{equ:csStar}) signal non-perturbative physics~\cite{Silverstein:2003hf} or non-trivial dynamics~\cite{Baumann:2011su}. 
Although finding $c_s > (c_s)_\star$ in future observations would disfavor the simplest deformation of slow-roll inflation, it would not constrain significantly other possible (well motivated) extensions.
This is the case because the EFT of inflation allows for stable hierarchies in which large interactions are possible without generating significant radiative corrections to the quadratic action.  These scenarios are not constrained by power spectrum measurements and are only probed by the bispectrum and/or higher $n$-point functions.  Similar thresholds of non-Gaussianity, $|\fnl^{\rm equil}|_\star \sim {\cal O}(1)$, exist in these cases. Moreover, the presence of additional light fields may also produce a signal above this threshold. This strongly motivates an experimental effort to improve the bounds on equilateral non-Gaussianity to the order-one level.
\vskip 4pt

Canonical slow-roll models may be thought of as the `Higgs mechanism' of inflation, in the sense that it is a weakly coupled completion of the EFT of inflation involving a fundamental scalar field. However, current observations have not yet ruled out the possibility that the UV completion of the EFT of inflation is characterized by strongly coupled dynamics, such as an analogue of compositeness in the electroweak sector, higher-derivative interactions or additional degrees of freedom. Unlike in particle physics, where higher energies can be explored by building more powerful accelerators,
 in cosmology we do not have direct access to the unitarity thresholds because observations are always performed at a fixed energy (corresponding to the freeze-out frequency during inflation). This means that sensitivity to physics at (or above) threshold can only be achieved by increasing the precision of the measurements. 
 In this paper we have shown
that observations of primordial B-modes together with the study of non-Gaussianity, in particular of the equilateral type, offer a unique window into the physics of the early universe and the nature of inflation. 
Finding that the primordial perturbations remain Gaussian beyond $|\fnl^{\rm equil}|_\star \simeq {\cal O}(1)$, or that measurements of the power spectra lead to $c_s > (c_s)_\star = 0.47$, would provide strong evidence for canonical slow-roll models. Conversely, a detection of non-Gaussianity above the threshold would open the road to physics beyond the slow-roll paradigm.

\subsubsection*{Acknowledgements}

We thank Valentin Assassi, Carlo Contaldi, Zhiqi Huang, Yi Wang, and Matias Zaldarriaga for helpful discussions.
D.B.~thanks Yi Wang for his patient assistance with the numerical analysis.
D.B.~gratefully acknowledges support from a Starting Grant of the European Research Council (ERC STG grant 279617).  R.A.P.~was partially supported by the German Science Foundation (DFG) within the Collaborative Research Center (SFB) 676 `Particles, Strings and the Early Universe', and by NSF grant AST-0807444 and DOE grant DE-FG02-90ER40542.

\newpage

\appendix

\section{Resumming Large Logarithms}\label{sec:flow}

In inflationary models with a non-trivial sound speed, 
scalars and tensors do not freeze out simultaneously and the tensor-to-scalar ratio depends non-trivially on the evolution of the Hubble parameter,  
\beq
r = 16 \varepsilon_1  c_s^{1+2\bar\varepsilon_1} \Big[1-{\cal C} \varepsilon_2 +(2-{\cal C})\delta_1 + \cdots\Big] \ , \label{equ:rRe}
\eeq
where
\beq
\bar \varepsilon_1 \equiv - \frac{\int \varepsilon_1(N)\hskip 1pt \d N}{\Delta N}\ , \qquad \Delta N \equiv N_t - N_s\ ,  \label{equ:EBAR}
\eeq
and the bracket in (\ref{equ:rRe}) contains higher-order terms in the slow-roll expansion. In general, the auxiliary parameter $\bar\varepsilon_1$ will be a function of the Hubble flow parameters $\varepsilon_n$.  We will assume a hierarchical structure and truncate the expansion at $\varepsilon_{n \geq 4}(N) \simeq 0$. This implies $\varepsilon_3 \approx const.$ and
\beq
\varepsilon_2(N) = \varepsilon_2(N_s)\, e^{\varepsilon_{3}(N-N_s)}\ .
\eeq
Our goal in this appendix is to obtain $\bar\varepsilon_1$  at leading order in $\varepsilon_{1,2}$ but to all orders in $\varepsilon_2\ln c_s$. In this way, \eqref{equ:rRe} becomes a resummation of the leading logarithms of the form $\bar\varepsilon_1 \ln c_s$, with $\bar\varepsilon_1$ incorporating all orders in $\varepsilon_2\ln c_s$.\footnote{Notice that for $\bar\varepsilon_1 \simeq \varepsilon_1 \approx const.$ the  expression in \eqref{equ:rRe} resums all the $\varepsilon_1\ln c_s$ terms.} For completeness, we will also discuss how to include $\varepsilon_3\ln c_s$ corrections. In what follows, we will suppress explicit reference to $N_s$, i.e.~all~the slow-roll parameters are understood to be evaluated at the freeze-out of the scalar modes.

\subsection[$\varepsilon_2 \ln c_s$]{${\boldsymbol{\varepsilon_2 \ln c_s}}$}

As we did throughout the main text, we consider $\varepsilon_3  \ll \varepsilon_{1,2}$ but keep all orders in $\varepsilon_{1,2}$. The solution for $\varepsilon_1$ then becomes
\beq
\varepsilon_1(N) = \varepsilon_1\, e^{\varepsilon_{2}(N-N_s)+ \tfrac{1}{2}\varepsilon_{2}\varepsilon_{3}(N-N_s)^2+\cdots}\ ,
\eeq
such that (\ref{equ:EBAR}) turns into
\beq
\bar \varepsilon_1 = \varepsilon_{1} \left[\frac{e^{\varepsilon_{2}\Delta N} - 1}{\varepsilon_{2}\Delta N} - e^{\varepsilon_{2}\Delta N} \frac{\varepsilon_{3} \Delta N}{2(\varepsilon_{2}\Delta N)^2}\Big\{2(1-e^{-\varepsilon_{2}\Delta N}) - 2\varepsilon_{2} \Delta N +(\varepsilon_{2}\Delta N)^2\Big\}\right]\ . \label{equ:EBAR2}
\eeq
Let us first look at the $\varepsilon_{1,2}$ dependence, as given by the leading term in (\ref{equ:EBAR2}). 
Then, using \eqref{equ:DNapp},
we have
\beq\label{equ:bareps1}
\bar\varepsilon_1 = \varepsilon_{1} \cdot \frac{e^{\varepsilon_{2}\Delta N(\bar\varepsilon_1)}-1}{\varepsilon_{2} \Delta N(\bar\varepsilon_1)}\ \ \to\ \ \bar\varepsilon_1 = \varepsilon_1 \cdot \frac{\left(1-c_s^{\varepsilon_{2}}\right)}{\varepsilon_1 \left(1-c_s^{\varepsilon_{2}}\right) - c_s^{\varepsilon_{2}}\varepsilon_2\ln c_s} \ .
\eeq
The resummation of the leading logarithms only requires
\beq\label{eq:be1}
\bar\varepsilon_1 \approx  \varepsilon_{1}\cdot \frac{1}{\varepsilon_2\ln c_s}\left(1-c_s^{-\varepsilon_{2}}\right) + \cdots\ .
\eeq
Substituting this into (\ref{equ:rRe}), we get
\begin{align}\label{equ:resum}
r &= 16 \varepsilon_1 c_s^{1+2 \varepsilon_1\cdot (1-c_s^{-\varepsilon_2})/(\varepsilon_2 \ln c_s)}  \Big[1-{\cal C} \varepsilon_2 +2(1-{\cal C})\delta_1+ \cdots\Big]\ . 
\end{align}
This is the result that we have implemented in our analysis in Section~\ref{sec:data}.

\subsection[$\varepsilon_3 \ln c_s$]{${\boldsymbol{\varepsilon_3 \ln c_s}}$}
The expression (\ref{equ:resum}) is valid to zeroth order in an expansion in $\varepsilon_3$.  
To compute the first-order correction we may proceed perturbatively.
Moreover, it is instructive to determine when the $\varepsilon_{1,2}\ln c_s$ terms become important. Keeping only the leading logarithms, the expression in \eqref{equ:EBAR2} turns into 
\beq
\bar \varepsilon_1 = \varepsilon_{1}\left[ 1 - \frac{1}{2}~\varepsilon_{2} \ln c_s\left(1 +\frac{1}{3}\varepsilon_3 \ln c_s\right) + {\cal O}\left(\varepsilon^2_n \ln c_s\right)\right] \label{equ:bareps1n}.
\eeq
 For $|\ln c_s| \leq 3.91$~({\rm Planck,~95\%CL}), the $\ln c_s$ terms produce an enhancement, but only become non-perturbative for somewhat large values of the slow-roll parameters: a perturbative treatment is justified provided $|\varepsilon_2| \ll 0.51$ and $|\varepsilon_3| \ll 0.75$. Values at the boundary of the perturbative regime are highly constrained by the running of the spectral index, $|\alpha_s| \lesssim 2 \times 10^{-2}$. The numerical analysis in \S\ref{sec:rob}, using $|\varepsilon_{1,2}| \lesssim 0.5$ and $|\varepsilon_3| \lesssim 0.1$, shows that it is self consistent to work within a slow-roll expansion, and to ignore the $\varepsilon_3$ correction in~\eqref{equ:bareps1n} on the boundaries of the 95\% confidence contours.

\newpage
\addcontentsline{toc}{section}{References}
\bibliographystyle{utphys}
\bibliography{Refs,books}

\end{document}